\documentclass[reprint,amsmath,amssymb,aps,pra]{revtex4-2}
\usepackage{graphicx}
\usepackage{dcolumn}
\usepackage{bm}
\usepackage{lmodern}
\usepackage{mathtools}
\usepackage{dsfont}
\usepackage{mathrsfs}
\usepackage{amsthm,amsfonts,mathtools}
\usepackage{mathbbol}
\usepackage{hyperref}
\usepackage{xcolor}
\usepackage{subfigure}

\newcommand{\rrho}{\hat{\boldsymbol{\rho}}}
\usepackage[normalem]{ulem}

\newcommand{\reals}{\mathds{R}}
\newcommand{\complex}{\mathds{C}}
\theoremstyle{plain}

\theoremstyle{definition}

\begin{document}

\preprint{APS/123-QED}

\title{Quantum-classical distance as a tool to design optimal chiral quantum walks}	

\author{Massimo Frigerio}
\email{Electronic address: massimo.frigerio@unimi.it}
\affiliation{Quantum Technology Lab $\&$ Applied Quantum Mechanics Group, Dipartimento di Fisica ``Aldo Pontremoli'', Universit\`a degli Studi di Milano, I-20133 Milano, Italy}
\affiliation{INFN, Sezione di Milano, I-20133 Milano, Italy}

\author{Claudia Benedetti}
\email{Electronic address: claudia.benedetti@unimi.it}
\affiliation{Quantum Technology Lab $\&$ Applied Quantum Mechanics Group, Dipartimento di Fisica ``Aldo Pontremoli'', Universit\`a degli Studi di Milano, I-20133 Milano, Italy}
\affiliation{INFN, Sezione di Milano, I-20133 Milano, Italy}

\author{Stefano Olivares}
\email{Electronic address: stefano.olivares@fisica.unimi.it}
\affiliation{Quantum Technology Lab $\&$ Applied Quantum Mechanics Group, Dipartimento di Fisica ``Aldo Pontremoli'',
Universit\`a degli 
Studi di Milano, I-20133 Milano, Italy}
\affiliation{INFN, Sezione di Milano, I-20133 Milano, Italy}

\author{Matteo G. A. Paris}
\email{Electronic address: matteo.paris@fisica.unimi.it}
\affiliation{Quantum Technology Lab $\&$ Applied Quantum Mechanics Group, Dipartimento di Fisica  ``Aldo Pontremoli'',
Universit\`a degli 
Studi di Milano, I-20133 Milano, Italy}
\affiliation{INFN, Sezione di Milano, I-20133 Milano, Italy}

\begin{abstract}
Continuous-time quantum walks (CTQWs) provide a valuable model for quantum transport, universal quantum computation and quantum spatial search, among others. Recently, the empowering role of new degrees of freedom in the Hamiltonian generator of CTQWs, which are the complex phases along the loops of the underlying graph, was acknowledged for its interest in optimizing or suppressing transport on specific topologies. We argue that the quantum-classical distance, a figure of merit which was introduced to capture the difference in dynamics between a CTQW and its classical, stochastic counterpart, guides the optimization of parameters of the Hamiltonian to achieve better quantum transport on cycle graphs and spatial search to the quantum speed limit without an oracle on complete graphs, the latter also implying fast uniform mixing. We compare the variations of this quantity with the 1-norm of coherence and the Inverse Participation Ratio, showing that the quantum-classical distance is linked to both, but in a topology-dependent relation, which is key to spot the most interesting quantum evolution in each case.
\end{abstract}

\maketitle
\section{Motivation and layout}

Continuous-time quantum walks (CTQW) are intensively studied as simplified models for a wide range of applications, spacing from quantum transport \cite{kendon2011perfect,mulken2011continuous,razzoli11,kulvelis15,tama16,tamascelli19,zatelli20}, e.g. excitonic transport in biochemical complexes involved in some instances of bacterial photosynthesis  \cite{kempe2003quantum,mohseni2008environment,Chin2010} , to universal quantum computation \cite{childs09comp} or specific quantum algorithms such as the spatial search \cite{childs04,portugal2013quantum,childs2014,chakraborty2016,Chakraborty20,paris21}. They are usually defined \cite{fahrigutman} on the same line of classical random walks (RW) on undirected, simple graphs, by promoting the graph Laplacian $\mathrm{L}$, which is the generator of the time evolution of an unweighted RW, to an Hamiltonian $\mathrm{H}$. 
Going from the classical transfer matrix to the quantum unitary evolution operator, together with the shift of focus from probabilities to amplitudes, make CTQW radically different from classical RW, with reliable prospects of achieving a quantum advantage in specific tasks \cite{chil02}. To mention a few, it is known that CTQW exhibit ballistic propagation of probability on lattices, contrary to the diffusive behavior of classical RW, and that they can solve  search problems in shorter times.

\par 
Fairly recently, however, it has been suggested that constraining the Hamiltonian of the CTQW to be the Laplacian of the graph is an unnecessary restriction, and richer phenomenology can be observed when the off-diagonal matrix elements of $\mathrm{H}$ are allowed to be generic complex phases, in compliance just with hermiticity: the resulting systems are called \emph{chiral} CTQW \cite{cqw1,cqw2}, because they exhibit asymmetry under time-reversal and directional bias in the propagation of probability. Further motivation for this generalization of CTQW has been provided in \cite{frigerio2021generalized}, where the most general correspondence between classical and quantum continuous-time random walks was also derived. Acknowledging this larger space of opportunities offered by quantum walks, which stems from the one-to-many nature of the step from classical to quantum, one is immediately faced with a new challenge: if assuming $\mathrm{H}= \mathrm{L}$ leads to a single choice of CTQW for a given unweighted graph, enlarging the focus to include chiral CTQW offers many free, real parameters that can be adjusted to optimize the quantum advantage in a specific task, at fixed graph topology. \\

\par 
In this paper, we show through a variety of relevant examples and an analytic discussion, that a previously introduced quantity \cite{GBP20}, the quantum-classical distance $\mathcal{D}_{QC}$, is a valuable tool to guide this optimization: it correctly captures the distinction between classical and quantum evolutions of random walks on graphs. Quite generally, quantum walks outperform classical random walks because of faster hitting, which is related to quantum transport and targeting, and faster mixing \cite{mooremixing,ahmadimixing}, when the walker spreads out towards a maximally coherent, uniform superposition of all sites. 
By maximizing the value of $\mathcal{D}_{QC}$ at short time scales over the free parameter's space, one is quickly directed towards chiral CTQWs with maximal advantage over the corresponding classical RW, either because of quick hitting or mixing \cite{ahmadimixing,mooremixing} depending on the graph's topology. 
Indeed, although from their very introduction in \cite{fahrigutman} the quantum advantage of CTQW has been linked with quantum coherence, it is clear that it cannot reduce to this sole quantum phenomenon, since the target state in a transport task on a graph has very little coherence, for example (although it \emph{does} play a role throughout the evolution).

\par 
After setting the notation for continuous-time classical and (chiral) quantum walks in Section \ref{sec:II},
we bring attention to the three dynamical quantities that will be employed throughout the article: the quantum-classical distance, the 1-norm of coherence and the Inverse Participation Ratio, which are all defined in Section \ref{sec:III} and accompanied by a comparison of their short-time expansions for generic chiral Hamiltonians. We then argue the effectiveness of $\mathcal{D}_{QC}$ in the identification of "optimal" phase configurations at fixed topology by looking at four very different and emblematic examples. We start with cycle graphs in Section \ref{sec:IV}, which allow for analytical expressions also in the presence of phases and constitute a testbed for our ideas. For odd cycles we show that the quantum-classical distance correctly signals the best phase for quantum transport and, conversely, spots the characteristic suppression of transport in even cycles for a resonant phase value, which is associated with a dip in the value of $\mathcal{D}_{QC}$.

\par 
Then in Section \ref{sec:V} we move to complete graphs, having maximal connectivity. After a preliminary exploration based on randomly generated Hamiltonians, we maximize the quantum-classical distance at short times and find a particular set of hermitian Hamiltonians which permit quantum search without oracle in a time which exactly achieves the optimal quantum speed limit, outperforming Grover's algorithm in the constant pre-factor and holding for any size of the complete graph. Interestingly, here $\mathcal{D}_{QC}$ is larger for evolutions which quickly lead to highly \emph{delocalized} states, in stark contrast with its behavior for cycles. This is consistent with the other typical behavior of quantum walks, which is \emph{fast mixing}. Indeed, the same optimal chiral evolution on complete graphs achieves uniform mixing, contrary to the non-chiral evolution, also with a quadratic speedup with respect to the standard protocol that involves hypercube graphs \cite{mooremixing}.

As a third case, in Section \ref{sec:VI} we examine quantum switches, constructed from a triangle graph with independent chains of sites attached to each vertex of the polygon. It is already known that, for resonant value of the sole free phase, this topology allows for directional quantum transport from one arm of the triangle to another, with minimal losses on the excluded, third arm. We show that the quantum-classical distance again spots the best value of the phase. Moreover, since these graphs are non-regular, an ambiguity about the diagonal entries of $\mathrm{H}$ arises, since the non-chiral association $\mathrm{H} = \mathrm{L}$ indirectly introduces a potential field landscape. We argue that setting all the diagonal phases of $\mathrm{H}$ to the same, arbitrary value, instead of the non-uniform connectivities of the different sites, is the unbiased choice and also the most efficient one for directional quantum transport. 

\par 
Finally, we tackle the cube graph in Section \ref{sec:VII}, on which perfect quantum transport from one vertex to the opposite one, with the Laplacian as generator, is known to happen. The quantum-classical distance here suggests that phases cannot improve this standard evolution. However, minimal values of $\mathcal{D}_{QC}$ are achieved in correspondence with evolutions that suppress transport at all times on half of the vertices of the cube.

\par
\section{Quantum and classical walks on graphs}
\label{sec:II}

Continuous-time quantum walks are traditionally introduced through an analogy with classical random walks on graphs. For the latter, one considers an undirected simple graph $\mathcal{G} =( V , E) $ of $N$ vertices, where $V$ is the set of vertices or \emph{sites}, and $E$ is the set of edges. To $\mathcal{G}$ corresponds an $N \times N$ symmetric matrix, namely the \emph{Laplacian matrix} of the graph, which is defined as $L = D - A$ where $A$ is the adjacency matrix of $\mathcal{G}$, such that $\left[ A \right]_{jk} = 1$ if there is an edge in $E$ connecting sites $j$ and $k$ and $0$ otherwise, including the diagonal elements, whereas $D$ is a diagonal matrix encoding the connectivities of each vertex, i.e. the number of edges departing from it. Since the sum of the rows and of the colums of $\mathrm{L}$ is zero, it generates a semigroup of bistochastic transformations $\mathcal{E}_{t} = e^{-t \mathrm{L} }$ for $t \in \reals^{+}$. Therefore, acting on a vector $\underline{p}_0 \in \reals^{N}$ of occupation probabilities for each site at the initial time, one has a continuous-time, autonomous stochastic process on the graph, which we will call a continuous-time (classical) random walk:
\begin{equation}
\label{eqCRWp}
    \underline{p} (t) \ \ := \ \ \mathcal{E}_{t} \left[ \underline{p} \right]\ \ := \ \ e^{-t \mathrm{L} } \underline{p}_{0}.
\end{equation}
Fahri and Gutman \cite{fahrigutman} noticed a suggestive similarity between Eq.(\ref{eqCRWp}) and the Schr\"{o}dinger equation for an $N$-level quantum system. Therefore, they proposed to \emph{define} a continuous-time quantum walk on the same graph $\mathcal{G}$ by promoting $\mathrm{L}$ to a quantum Hamiltonian matrix $\mathrm{H} = \mathrm{L}$ acting on an $N$-dimensional complex Hilbert space $\mathcal{H} \sim \mathds{C}^{N}$ which has a preferred basis $\{ \vert j \rangle \}_{j= 1, ... N}$ whose elements describe localized quantum states on the sites of the graph. The quantum evolution of any initial state $\vert \psi_{0} \rangle \in \mathcal{H}$ now comes automatically \footnote{ Here and in the following we fix $\hbar = 1$. The units of energies are reabsorbed in the time scale so as to conveniently keep everything dimensionless.}:
\begin{equation}
    \vert \psi(t) \rangle \ = \ U_{t} \vert \psi_{0} \rangle \ = \ e^{-i t \hat{H} } \vert \psi_{0} \rangle.
\end{equation}
Recently it was pointed out that imposing $\mathrm{H} = \mathrm{L}$ is too restrictive on the quantum side, since a generic Hamiltonian compatible with the graph topology does not need to be a real matrix, but just a Hermitian one. We will focus on \emph{unweighted} graphs, such that the  non-zero elements of the adjacency matrix $A$ are equal to $1$ for all the links. Therefore, the most general $\mathrm{H}$ has off-diagonal entries:
\begin{equation}
   \left[ \mathrm{H} \right]_{kj} \ \ =  \ \ e^{ i \varphi_{kj}} \ \ \ \ ( j \neq k )
\end{equation}
if sites $j$ and $k$ are linked in $\mathcal{G}$, and zero otherwise. Here $\varphi_{kj} \in [0, 2 \pi)$ is a phase depending on the link and $\varphi_{kj} = - \varphi_{jk}$ is required by Hermiticity of $\mathrm{H}$. On the other hand, the diagonal elements of $\mathrm{H}$ must be real numbers, but they do not need to satisfy any further constraint. For regular graphs, the simplest and most unbiased choice is to assume  $\left[ \mathrm{H} \right]_{jj} = d$ for all $j = 1,...,N$, so that they are all equal and they can be discarded as they generate an overall, unobservable phase shift on the whole Hilbert space. For non-regular graphs, instead, the choice is less obvious and we shall return to this point in a later section (see also \cite{wong2016laplacian}). Continuous-time quantum walks whose Hamiltonian is a generic Hermitian matrix compatible with the graph topology are also called \emph{chiral} CTQW, to distinguish them from the most standard case with $\mathrm{H} = \mathrm{L}$ (or $\mathrm{H} = \mathrm{A}$), which is by far the most studied in the literature as of now. 
\newline

In this work, we are interested in optimizing the additional degrees of freedom of a chiral CTQW on a given graph to achieve the maximal advantage over the corresponding, unique classical RW. It is thus convenient to formulate both evolutions in the Hilbert space formalism to facilitate the comparison. To do this, we first remark that quantum coherent states in the site basis do not exist at the classical level, therefore we shall assume a localize state $\vert j \rangle$ with $j = 1,...,N$ as the initial condition for both the quantum and the classical evolution, to make a fair comparison. Then the classical evolution can be defined by the standard embedding of a classical probability onto a Hilbert space with respect to a preferred basis, that is:
\begin{equation}
\label{eq:CLev}
    \mathcal{E}_{t} \big[ \vert j \rangle \langle j \vert \big]  = \sum_{k=1}^{N}   \langle k \vert e^{-t \mathrm{L}} \vert j \rangle   \vert k \rangle \langle k \vert  = \sum_{k=1}^{N} p_{kj}(t) \vert k \rangle \langle k \vert
\end{equation}
which ensures that the classically evolved state is always incoherent in the localized basis. Also the quantum evolution can be rewritten in a similar way:
\begin{equation}
\label{eq:Qev1}
    \mathcal{U}_{t} \big[ \vert j \rangle \langle j \vert \big] \  = \ e^{-i \hat{H} t} \vert j \rangle \langle j \vert e^{ i \hat{H} t } \  =  \ \vert \psi_{j} (t) \rangle \langle \psi_{j} (t) \vert
\end{equation}
where:
\begin{align}
\label{eq:Qev2}
    \vert \psi_{j}(t) \rangle \ \ & = \ \ \sum_{k=1}^{N} \alpha_{kj}(t) \vert k \rangle \\
    \alpha_{kj}(t) \ \ &= \ \ \langle k \vert e^{-i\hat{H} t} \vert j \rangle \ \ = \ \ \left[  e^{-i \mathrm{H} t} \right]_{kj}.
\end{align}

\subsection{Chiral quantum walks and gauge invariance of transition probabilities}
By a diagonal, unitary change of basis, which is a local phase transformation, many of the phases of $\mathrm{H}$ can be set to $0$. This is particularly useful when considering site-to-site transition probabilities, as it is often the case when dealing with quantum walks. Indeed, quantities such as $P_{j \to k} (t) = \vert \langle k \vert e^{- i H t} \vert j \rangle \vert^{2}$ and functions thereof do not depend on phases which can be removed by those local phase transformations. These are effectively \emph{gauge transformations} and, at least for some regular topologies of the underlying graphs, the change in the Hamiltonian's phases can be accounted for by a gauge transformation of the \emph{gauge connection field}. Since to each edge of the graph can correspond a phase in $\mathrm{H}$ and the phase of the localized state on each vertex can be modified independently, the maximal number of phase degrees of freedom that affect the transition probabilities $P_{j \to k}(t)$ is $|E|-N + 1 = 1 - \chi(\mathcal{G}) $, where $|E|$ is the number of edges of the graph, whereas $\chi(\mathcal{G})$ is the Euler characteristic for generic graphs, including non-planar ones.
For planar graphs, this number is precisely the number of loops, while for non planar ones the number of loops is not obvious to define. In particular, phases cannot affect transition probabilities on tree graphs, while cycle graphs all have just a single relevant phase, which can be distributed over each link or concentrated on a single one. However, given a particular choice for the Hermitian matrix $\mathrm{H}$, it is not trivial to determine whether it is equivalent to a real, symmetric Hamiltonian or not. It has been shown \cite{TRBIAM21} that this is the case if and only if the product of phase factors along each directed, simple, closed path on $\mathcal{G}$ is $1$. 
\section{Quantum-classical distance}
\label{sec:III}
\label{sec:2}
By combining Eq.(\ref{eq:CLev}) and Eq.(\ref{eq:Qev1},\ref{eq:Qev2}), it is possible to define a time-dependent quantity which compares the quantum evolution to the classical one, when both start in the same, localized initial state $\rrho^{0}_{j} = \vert j \rangle \langle j \vert$. This quantity is the \emph{quantum-classical distance} \cite{GBP20}:
\begin{equation}
    \mathcal{D}_{QC}^{j} (t)  \ \ := \ \  1 - \sum_{k=1}^{N} p_{kj} (t) \vert \langle k \vert \psi_{j} (t) \rangle \vert^{2} .
\end{equation}
It is still true, by the same token as in \cite{GBP20}, that the minimum value of the quantum-classical distances over all possible \emph{classical} initial states can be attained by a single localized state, at any given time. Therefore we can still define a global quantum-classical distance by \footnote{$V$ is the ordered set of vertices of the graph.}:
\begin{equation}
 \mathcal{D}_{QC} (t) \ \ :=  \ \ \max_{j \in V}  \left\{ \mathcal{D}_{QC}^{j} (t) \right\}
\end{equation}
which is independent of the initial state, but only depends upon the graph (specified by $\mathrm{L}$) and the choice of $\mathrm{H}$ compatible with the topology.\\

Checking for a few graphs with different topologies and number of vertices, the overall behaviour of this quantity in time seems very similar to the simpler scenario where $\mathrm{L}$ is a Laplacian and the straightforward identification $\mathrm{H} = \mathrm{L}$ can be applied. In particular, the asymptotic value in the long time limit takes the same expression for any connected graph of size $N$ and it is given by:
\begin{equation} 
\mathcal{D}_{QC}^{j} (t \gg 1) \simeq 1 - 1/N.
\end{equation}

Because of this saturation effect of $\mathcal{D}_{QC}$ at large times, fluctuations of the quantum-classical distance can be significant only if they happen before the classical evolution gets close to equilibrium. This is a valuable property: indeed, since the quantum evolution is aperiodic on general graphs, wild behaviors can occur at sufficiently long times and it is therefore mandatory, from a physical point of view, to impose a cutoff on the relevant time scale to perform the required task. A natural choice is precisely the time scale of the classical evolution, especially if a quantum advantage is sought.

\par
\subsection{Quantum-classical distance at short times}
 To understand the behavior of the quantum-classical distance at short times, instead, we may expand $p_{kj} (t)$ and $ \vert \langle k \vert \psi_{j} (t) \rangle \vert^{2}$ up to second order in $t$ and obtain
\begin{align}
    \mathcal{D}_{QC}^{j} (t)  =& \, t \left( \left[ \mathrm{H}^{2} \right]_{jj}  - [ \mathrm{H}]_{jj}^{2}  \right) \notag \\ 
    & - t^2  \Bigg(- \left[\mathrm{H}^{2} \right]_{jj}  + [ \mathrm{H}]_{jj}^{2} + \frac12 
    \left[\mathrm{H}^{2} \right]_{jj}^{2} \notag \\ 
    & \; - \left[\mathrm{H}^{2} \right]_{jj} [\mathrm{H}]_{jj}^{2} + \frac12 \sum_s \left| \mathrm{H}_{js}\right|^{4}
     \Bigg) \ + \ O(t^3)
\end{align} 
If we focus on unweighted graphs, which means that we assume $\vert  \mathrm{H}_{jk} \vert$ to be either $1$ or $0$ depending on whether vertices $j$ and $k$ are connected by an edge or not, then we can considerably simplify the expression above. Indeed, we find that:
\begin{equation}
    \left[ \mathrm{H}^{2} \right]_{jj}  - [ \mathrm{H}]_{jj}^{2}  \  =  \ d_{j} \  = \ \mathrm{L}_{jj}
\end{equation}
where $d_{j}$ is the connectivity of vertex $j$ in the graph. Notice also that:
\begin{align}
 \frac12 
   & \left[\mathrm{H}^{2} \right]_{jj}^{2} - \left[ \mathrm{H}^{2} \right]_{jj} [\mathrm{H}]_{jj}^{2} + \frac12 \sum_s \left| \mathrm{H}_{js}\right|^{4} \ = \\
  & = \frac12 \left( \left[ \mathrm{H}^{2} \right]_{jj}  - [ \mathrm{H}]_{jj}^{2} \right)^{2} + \frac12 \sum_{s \neq j}  \left| \mathrm{H}_{js}\right|^{4} \ = \\
  & = \frac{d^{2}_{j} + d_{j}}{2} 
\end{align}
where in the last step we used the fact that $\sum_{s \neq j} \left| \mathrm{H}_{js}\right|^{4}$ gets a contribution of $1$ for any $s$ connected to $j$ and $0$ otherwise. Therefore, the short-time expansion of the quantum-classical distance to second order is:
\begin{equation}
\label{eq:DQCshortT}
    \mathcal{D}_{QC}^{j} (t)  \ = \ d_{j} t \ - \  d_{j} (d_{j} - 1) \frac{t^{2}}{2} \ + \ O(t^3).
\end{equation}
It is relevant to stress that the coefficients of this expansion are dictated just by the connectivity of the starting vertex and they are not influenced by any other parameter involved in $\mathrm{H}$: in particular, \emph{neither the on-site average energies nor the phases of the off-diagonal entries of $\mathrm{H}$ affect the short-time behavior of the quantum-classical distance}.

\subsection{1-norm of Coherence and Inverse Participation Ratio}
In order to understand what type of properties of the quantum walk contributes to the quantum-classical distance, and how $\mathcal{D}_{QC}$ relates to some figure of merits that are considered useful for particular tasks, we now seek the short-time expansion of other two quantities, the 1-norm of coherence and the inverse participation ratio (IPR), to be later compared with Eq.(\ref{eq:DQCshortT}).

\par
Let us start with the 1-norm of coherence. With respect to the localized basis, the coherence of the pure state $\vert \psi_{j}(t) \rangle$ given by Eq.(\ref{eq:Qev1}) can be written as \cite{GBP20}:
\begin{equation}
    \mathcal{C}_{j} (t) \ \ = \ \ \left( \sum_{k} \vert \alpha_{kj} (t) \vert \right)^{2} - 1 
\end{equation}
with $\alpha_{jk}(t)$ defined according to Eq.(\ref{eq:Qev2}). Here and in the following, it will be helpful to start from the fourth-order short time expansion of $\vert \alpha_{kj} (t) \vert^{2}$:
\begin{align}
\label{eq:alfa2smallt}
    \vert \alpha_{kj}(t) \vert^{2}  &= \ \ \delta_{jk} \ - \ t^{2} \left( \delta_{jk} \left[ \mathrm{H}^{2} \right]_{jj} - \vert \mathrm{H}_{jk} \vert^{2} \right) \ + \\ 
    & \  + \  t^3 \mathrm{Im} \left[ \mathrm{H}_{jk} \left[ \mathrm{H}^{2} \right]_{kj} \right] \ + \notag \\
    & + \ t^4 \left(  \frac{1}{12} \delta_{jk} \left[ \mathrm{H}^{4} \right]_{jj} - \frac{1}{3} \mathrm{Re} \left[ \mathrm{H}_{jk} \left[ \mathrm{H}^{3} \right]_{kj} \right] + \right. \notag \\
    &  \ + \ \left. \frac14 \vert \left[ \mathrm{H}^{2} \right]_{jk} \vert^{2}          \right) \ + \ O(t^5) . \notag
\end{align}
From this, it follows that:
\begin{align}
    & \sum_{k}  \vert \alpha_{kj} (t) \vert \ = \ \sqrt{ 1 -  d_{j} t^{2} + O(t^4)} \ + \\
    &  + \ t \sum_{k \neq j}  \vert \mathrm{H}_{jk} \vert \sqrt{ 1 - t\mathrm{Im} \left[ \mathrm{H}_{jk} \left[ \mathrm{H}^{2} \right]_{kj} \right] + O(t^2)}  \ + \notag \\
    & + \  \frac{t^2}{2} \sum_{k \neq j , \mathrm{H}_{jk} = 0}  \vert \left[ \mathrm{H}^{2} \right]_{jk} \vert \ + \ O(t^3) \notag
\end{align}
where we splitted the initial sum over $k$ in a first term with $k=j$, a second sum with $k \neq j$ but which gets contributions only for terms having $\mathrm{H}_{jk} \neq 0$, and a final sum that takes into account the last terms with $k \neq j$ and $\mathrm{H}_{jk} = 0$. Expanding the square roots and collecting powers of $t$, one obtains:
\begin{align}
    & \sum_{k}  \vert \alpha_{kj} (t) \vert \ = \  1 + t d_{j} - \frac12  t^{2} \Bigg( d_{j} \ +   \\ 
    & + \  \sum_{k \neq j}\mathrm{Im} \left[ \mathrm{H}_{jk} \left[ \mathrm{H}^{2} \right]_{kj} \right] + \sum_{\substack{k \neq j \\ \mathrm{H}_{jk} = 0}}  \vert \left[ \mathrm{H}^{2} \right]_{jk} \vert \Bigg)  + O(t^3). \notag
\end{align}
Now notice that:
\begin{align*}  \
\sum_{k \neq j}\mathrm{Im} \left[ \mathrm{H}_{jk} \left[ \mathrm{H}^{2} \right]_{kj} \right] \ &= \ \sum_{k}\mathrm{Im} \left[ \mathrm{H}_{jk} \left[ \mathrm{H}^{2} \right]_{kj} \right]    \ = \\
&= \ \mathrm{Im} \left[ \mathrm{H}^{3} \right]_{jj}  \ = \ 0 .
\end{align*}
Overall, for coherence we find:
\begin{align}
\mathcal{C}_{j}(t) & \ = \ \ 2 d_{j} t \ \  + \\
& + \ \Big[ d_{j} ( d_{j} - 1) - \sum_{\substack{k \neq j \\ \mathrm{H}_{jk} = 0 }}  \vert \left[ \mathrm{H}^{2} \right]_{jk} \vert  \Big] t^{2} \ + O(t^3). \notag
\end{align}
One can appreciate the fact that coherence has a very similar structure to quantum-classical distance at short times, but it is affected by other degrees of freedom of $\mathrm{H}$. Indeed, the term $\sum_{\substack{k \neq j \\ \mathrm{H}_{jk} = 0}}  \vert \left[ \mathrm{H}^{2} \right]_{jk} \vert$ has a simple graphical interpretation: $ \vert \left[ \mathrm{H}^{2} \right]_{jk} \vert$ for $k \neq j$ and for $\mathrm{H}_{jk} = 0$  is a sum over all possible paths of length $2$ connecting vertex $j$ with a vertex $k$ at distance $2$ from $j$, where each path is weighted by the product of the phases of its two links in the direction from $j$ to $k$. Since this is the modulus of a sum of phases, interference may happen. Therefore $\mathcal{C}_{j}$ can get a phase-dependent second order contribution in $t$ only if the graph contains a 4-cycle passing through $j$ and bearing a non-zero overall phase, so that there will be at least two paths joining the same $k$ with $j$ and weighted by different phase factors. \\

To quantify the spreading in time of the QW from the initial vertex, a useful figure of merit is provided by the {inverse participation ratio} :
\begin{equation}
    \mathcal{I}_{j} (t) \ \ = \ \ \sum_{k} \vert \alpha_{kj}(t) \vert^{4} 
\end{equation}
which decreases while the QW spreads over the vertices of the graph. The second-order small time expansion of $\mathcal{I}_{j}(t)$ is  derived from Eq.(\ref{eq:alfa2smallt}):
\begin{equation}
    \mathcal{I}_{j}(t) \ \ = \ \ 1 - 2 d_{j} t^{2} \ + \ O(t^4).
    \label{ipr}
\end{equation}
 The above expression is clearly related to coherence and $\mathrm{D}^{j}_{QC}$, but unlike for coherence, the IPR is determined just by the connectivity of the starting vertex at least up to second order in $t$. 
 
The conclusion we can draw is that the short-time expansion of the quantum-classical distance correlates with coherence and IPR, but in a way that depends non-trivially upon the topology; indeed, $\mathcal{D}_{QC} (t)$ compares the classical and quantum \emph{probability distributions} at a given time and these can be similar irrespective of the coherences. In the following, we shall argue that this is in fact a merit of $\mathcal{D}_{QC}$.

\section{Cycle graphs}
\label{sec:IV}
The simplest example to investigate the role of phases in the Hamiltonian of CTQW for quantum transport is provided by cycle graphs. As discussed previously, by gauge invariance we can write the generic Hamiltonian $\mathrm{H}$ for a chiral CTQW on a cycle graph as follows:
\begin{equation}
    \mathrm{H} \ \ = \ \ \left(  
    \begin{array}{cccccc}  d_{1} & e^{i \theta} & 0 & 0 & \ldots & e^{-i \theta} \\
    e^{-i\theta} & d_{2} & e^{i \theta} & 0 & \ldots & 0 \\
    0 & e^{-i\theta} & d_{3} & e^{i\theta} & \ldots &  0 \\
    \vdots & \vdots & & & \vdots & \vdots \\
    e^{i \theta} & 0 & 0 & \ldots  & e^{-i\theta} & d_{N} \end{array}
    \right)
\end{equation}
where $\theta \in [0, 2 \pi)$ is the only relevant phase, which has been distributed equally over all links for convenience, while $d_{1},...,d_{N} \in \reals^{+}$. In this work we focus on the role of phases, hence we will take $d_{1} = ... = d_{N} = d$ and further impose $d=0$ without loss of generality.
The eigenvectors of $\mathrm{H}$ will still be Bloch states:
\begin{equation}
    \vert \lambda_{j} \rangle  \ \ := \ \ \frac{1}{\sqrt{N}} \sum_{k=1}^{N} e^{i\frac{2 \pi j k}{N}} \vert k \rangle
\end{equation}
while the eigenvalues will be shifted according to:
\begin{equation}
    \lambda_{j} \ \ := \ \ 2 \cos \left( \theta + \frac{2 \pi j}{N} \right)
\end{equation}
where $j = 1, ..., N$. 
Notice that, although for $\theta = 0$ the spectrum is doubly degenerate, for almost every $\theta \neq 0$ the degeneracy is lifted. Moreover, despite $\theta \in [ 0, 2 \pi )$, one can always choose a representative phase in the reduced scheme, that is $\theta \in [ 0, \frac{2 \pi}{N} ) $, much like the first Brillouin zone for crystal momentum, without affecting transition probabilities between sites (see below).  When $N$ is large, the effect of $\theta$ on the spectrum is clearly minor. The transition probabilities associated with this chiral quantum walk on a cycle are:
\begin{align}
    P_{j \to k}(t) \  &:
    = \frac{1}{N^2} \Big\vert \sum_{s=1}^{N} e^{2 i  \left[ \frac{ \pi (k-j) s}{N}  - t \cos \left( \theta + \frac{2 \pi s}{N} \right) \right] } \Big\vert^{2}. \label{eq:pjkring}
\end{align}
We again stress the fact that, by virtue of gauge invariance, all the quantities that can be computed from these probabilities, assuming a localized initial state as before, would be the same for a more general $\mathrm{H}$ on the ring of the form:
\begin{equation}
\label{eq:Hring}
    \mathrm{H} \ \ = \ \ \left(  
    \begin{array}{cccccc}  d& e^{i \phi_{1}} & 0 & 0 & \ldots & e^{-i \phi_{N}} \\
    e^{-i\phi_{1}} & d & e^{i \phi_{2}} & 0 & \ldots & 0 \\
    0 & e^{-i\phi_{2}} & d & e^{i \phi_{3}} & \ldots &  0 \\
    \vdots & \vdots & & & \vdots & \vdots \\
    e^{i \phi_{N}} & 0 & 0 & \ldots  & e^{-i \phi_{N-1} } & d \end{array}
    \right) \notag
\end{equation}
by simply putting $\theta = \frac{1}{N} \sum_{j=1}^{N} \phi_{j}$ in Eq.(\ref{eq:pjkring}).

In the large ring limit, $N \to \infty$, since the single relevant phase $\theta$ can be concentrated as $N \theta$  on any single link at will by leveraging gauge invariance, the transition probabilities will be the same as for the standard continuous-time quantum walk on a ring:
\begin{equation}
    \lim_{N \to \infty}   P_{j \to k}(t) \ \ = \ \ \vert J_{ \vert k - j \vert} (2t) \vert^{2} 
\end{equation}
where $J_{n}(x)$ is the $n$-th Bessel function. The same result can also be derived directly from Eq.(\ref{eq:pjkring}). This also implies that the effect of phases on the quantum evolution starting on any fixed vertex will show up only when the probability will have reached the opposite side of the cycle, so that phase-dependent interference can happen. Since the initial propagation on the cycle is known to be ballistic with velocity $2$, in order to reach the opposite side of the cycle the walker will take a time approximately given by $\tau = N/4$, which therefore is also the time it takes for $\theta \neq 0$ to have some appreciable effect on the on-site probabilities. However, this prediction  can be inaccurate for small rings, because of the non-negligible tails of the ballistic wavefronts. \\

\begin{figure*}[!ht]
\centering
\subfigure{\includegraphics[scale=0.57]{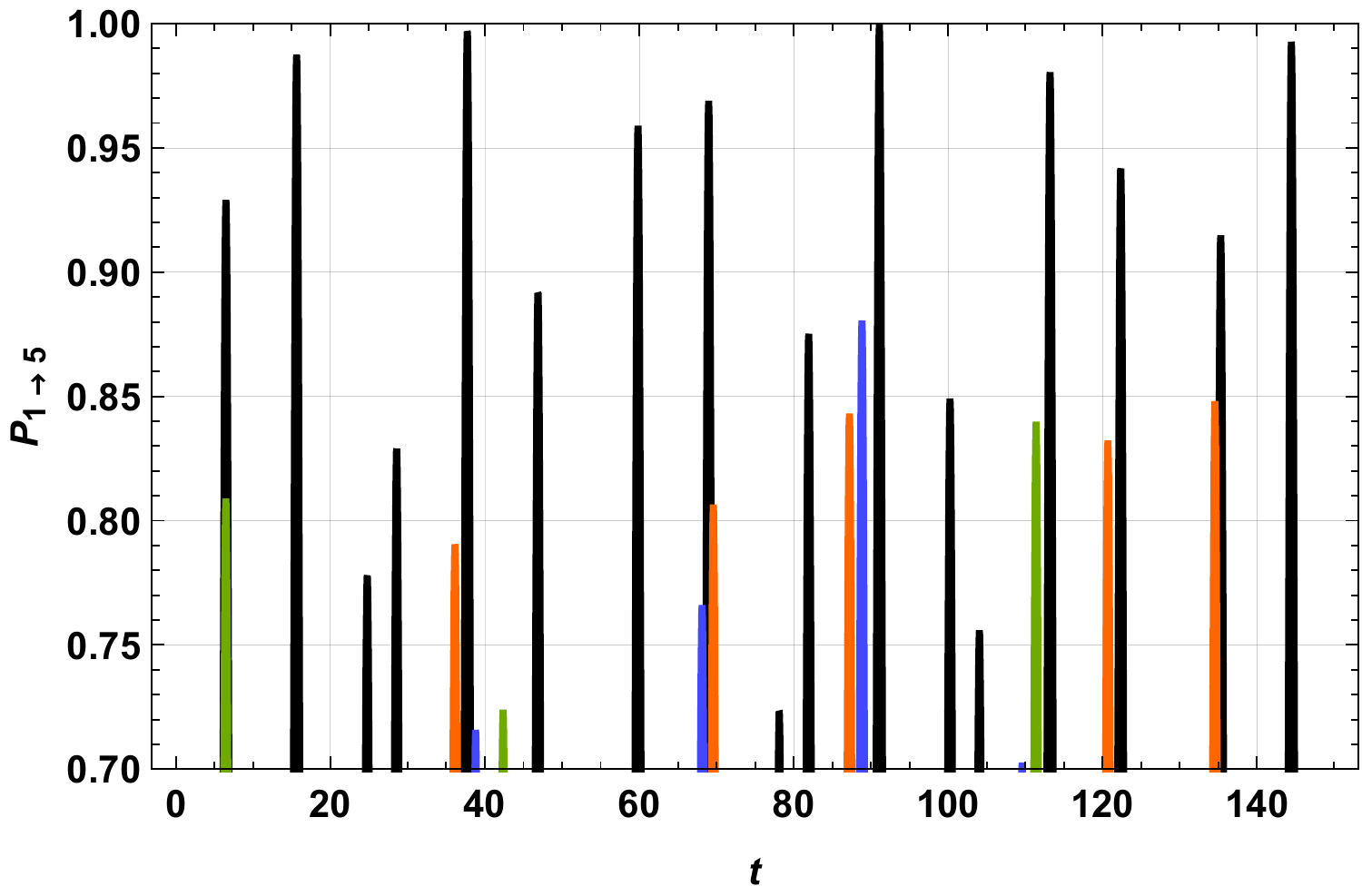}}\hfill
\subfigure{\includegraphics[scale=0.59]{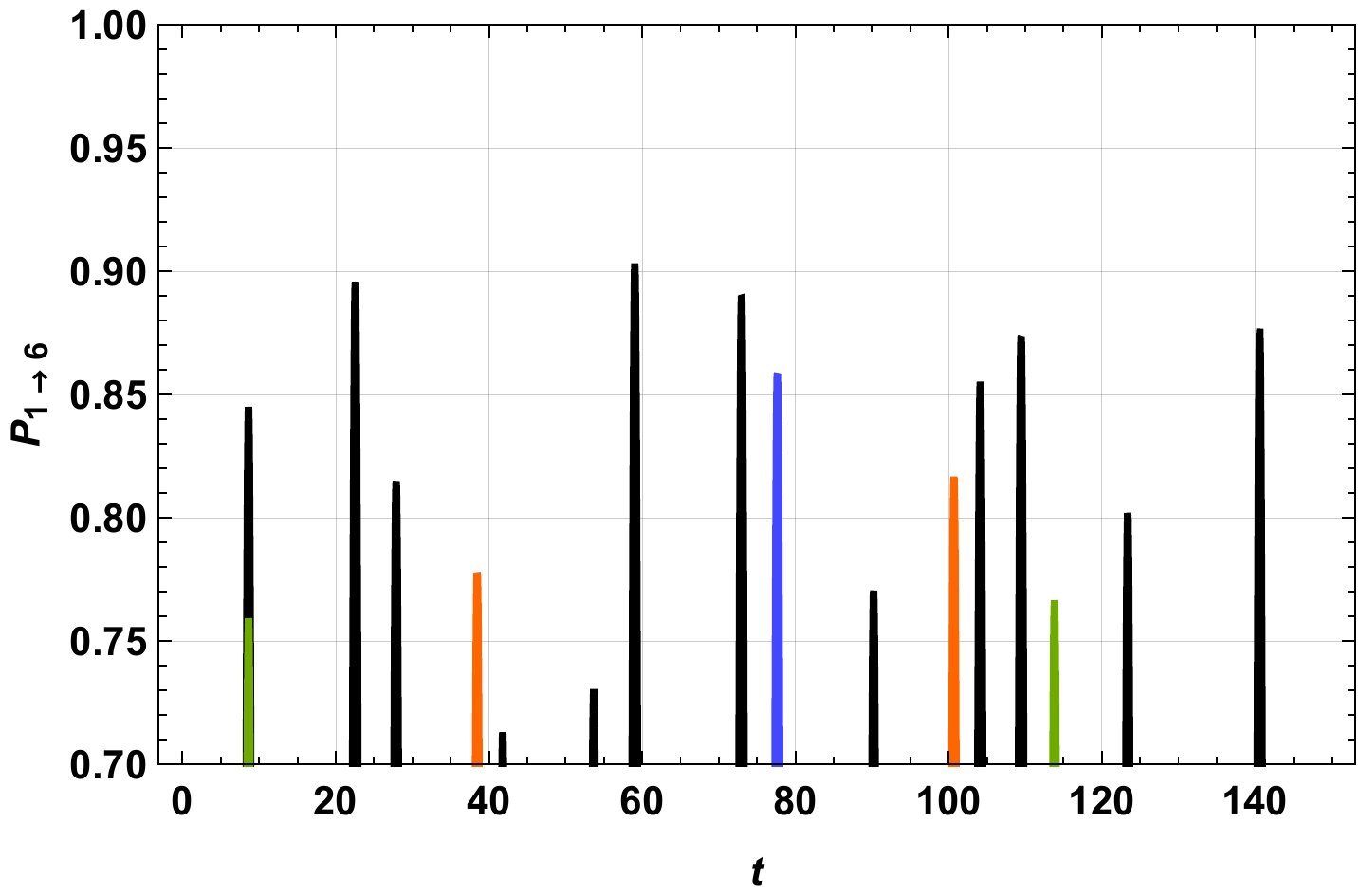}}
\caption{\small Transition probability from initial site $1$ to site $5$ of a 8-cycle (left) and to site $6$ of a 10-cycle (right) as functions of time and for different values of the phases. The non-chiral $\theta = 0$ case is plotted in black. For the 8-cycle the other phases are $\theta = 0.04$ (green), $\theta = 0.13$ (orange) and $\theta = 0.23$ (light blue). For the 10-cycle we chose $\theta = 0.027$ (green), $\theta = 0.13$ (orange) and $\theta = 0.28$ (light blue). Notice that the probability axis plot range is restricted to $[0.7,1]$ because it is the most informative region for quantum transport.}
\label{fig:1} 
\end{figure*}

To understand the influence of phases on quantum transport on the cycle after this minimal time, let us first consider the standard $\theta = 0$ case. Here, starting with an initially localized state $\vert j \rangle$, all sites that are symmetrical with respect to the initial vertex must have the same occupation probability at all times, since the symmetry is preserved by the non-chiral Hamiltonian. In particular, the probability of finding the walker on each site except the starting one (and the one opposite to it in the case of even cycles) cannot exceed $\frac12$ at any time. 
When the phase is added and $N$ is even, for any $\theta$ one can show that this symmetry is preserved. Indeed, even-cycles belong to the family of bipartite graphs, defined as those graphs whose vertices can be divided into two sets such that two vertices in the same set are never connected by an edge. For such graphs, one can apply a gauge transformation that flips the sign of all the basis elements of one set, while leaving the other set untouched, thereby implementing the transformation $\mathrm{H} \to - \mathrm{H}$, equivalent to time reversal, without affecting the transition probabilities. In particular, for even cycles this implies that the probability of landing at site $k$ or at its symmetric counterpart $N - k + 1$ is the same at all times also for $\theta \neq 0$.
Therefore, any chiral quantum walk on an even cycle with $N$ sites starting at site $j$ can never be localized with probability greater than $\frac12$ on any site different from the initial one and its opposite at $k = j + \frac{N}{2}$. As for the opposite site, which can be interesting as a target for transport, the situation is less clear. At fixed time, the optimal phase to maximize the probability $P_{j \to j+N/2}(t)$ can be different from zero. However, it seems that the highest peaks of $P_{j \to j+N/2}(t)$  for a wide range of times are always attained for $\theta = 0$. This can be appreciated in Fig.\ref{fig:1}, where we plotted the transition probabilities between opposite sites vs. time for cycles of $8$ and $10$ sites and for different values of $\theta$. We restricted the plotting region to probability values in $[0.7,1]$ because we considered $0.7$ as a lower bound to the acceptable fidelity of transport. Again, transition probabilities between other pairs of sites are neglected because they are always upper bounded by $\frac12$.

These observations suggest that chiral CTQWs provide no advantage for optimal quantum transport on even cycles over their non-chiral counterparts. Nevertheless, we highlight the fact that $P_{j \to j + N/2}(t)$ can be completely suppressed at all times by choosing $\theta = \pi/N$ (or, equivalently, a phase of $\pi$ on a single, generic link), a phenomenon that, together with the unbroken reflection symmetry for generic $\theta$, has already been attributed to the fact that even cycles are \emph{bipartite} graphs. We notice that, in the context of excitonic transport in biochemical complexes, this sets a possible prediction to be contrasted with observations, since finding a ring-like structure with an even number of units could exclude that phases play a role.\\

\begin{figure*}[!ht]
\centering
\subfigure[a]{\includegraphics[scale=0.59]{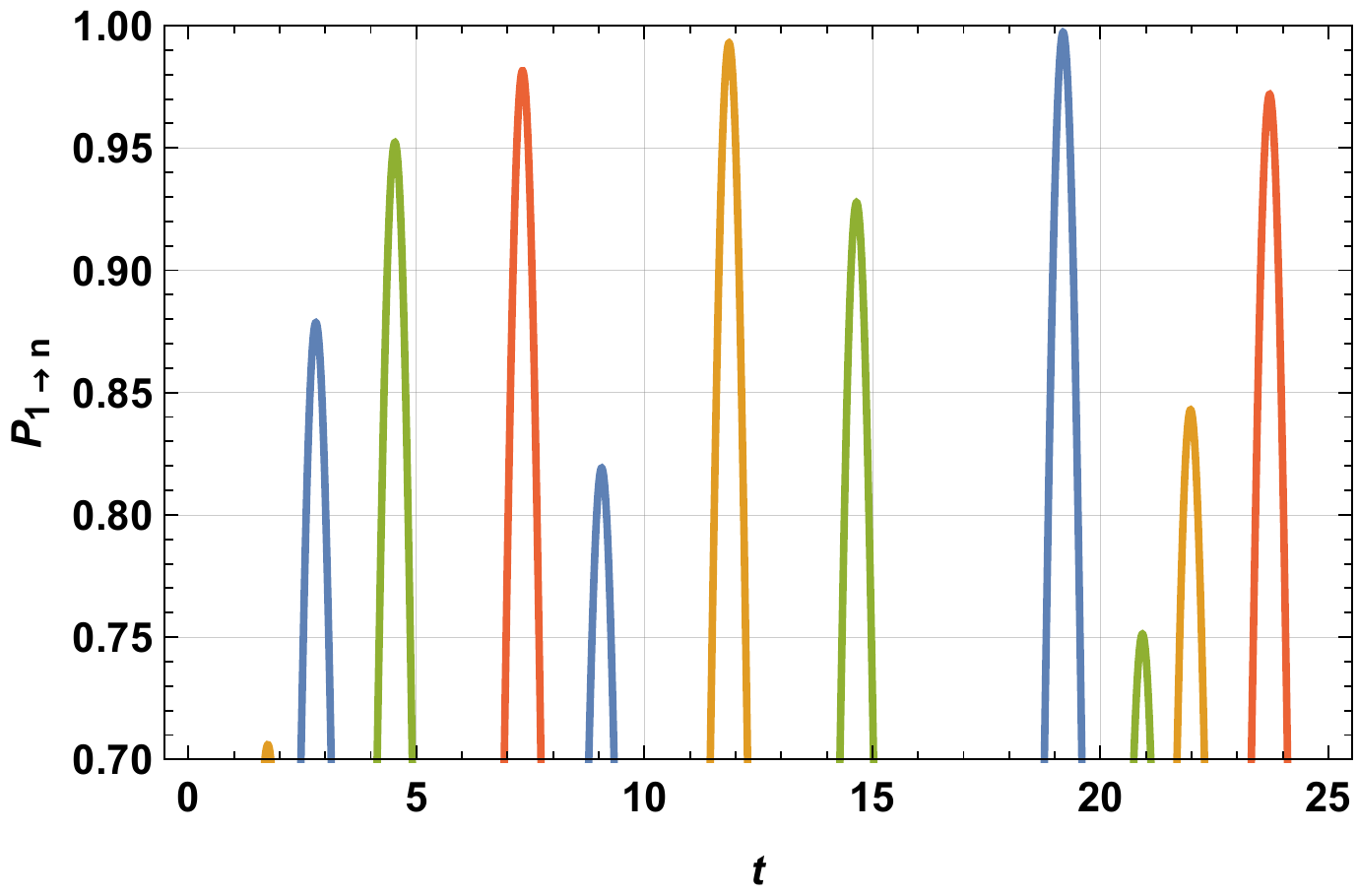}}\hfill
\subfigure[b]{\includegraphics[scale=0.59]{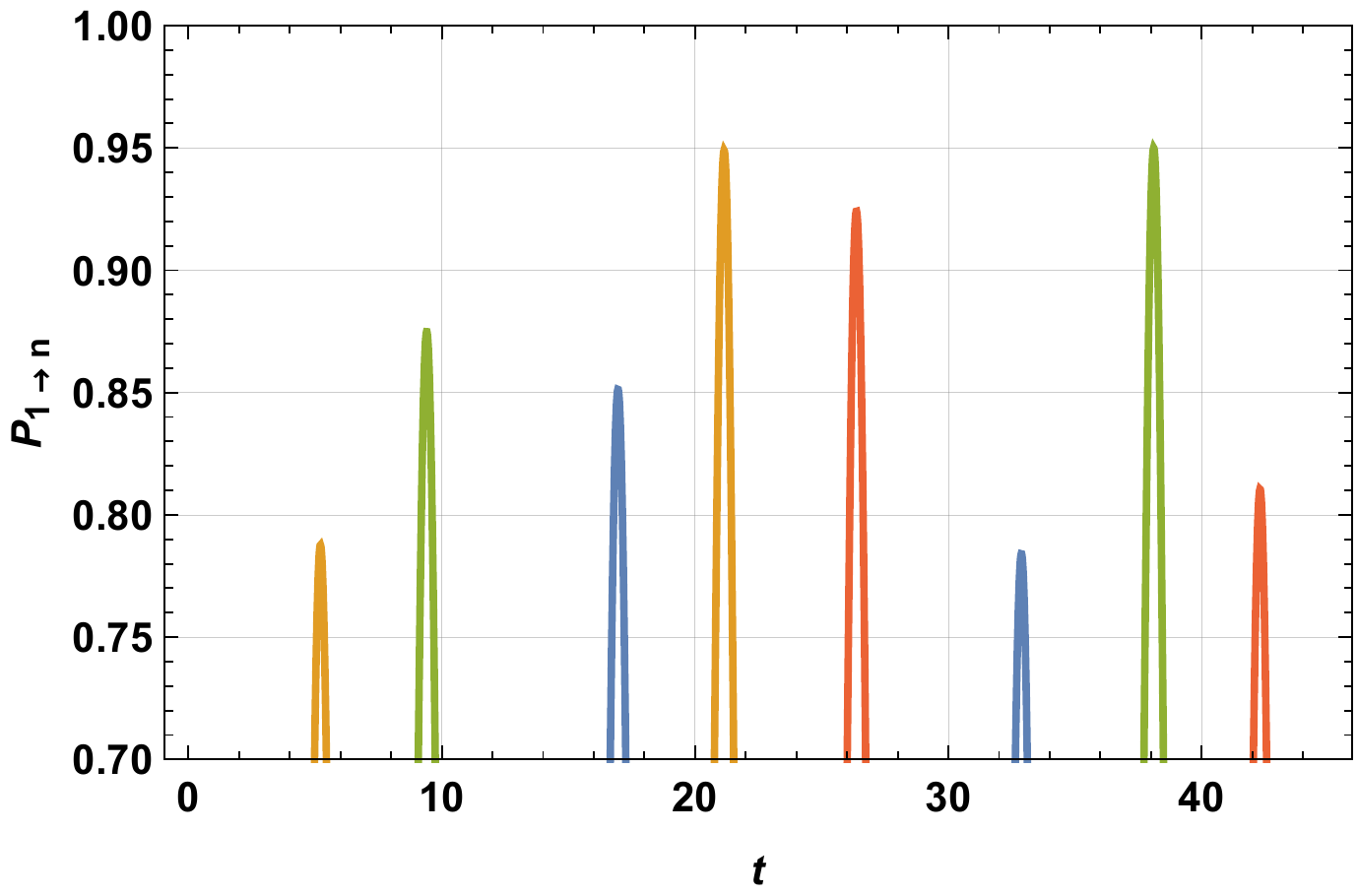}}
\subfigure[c]{\includegraphics[scale=0.59]{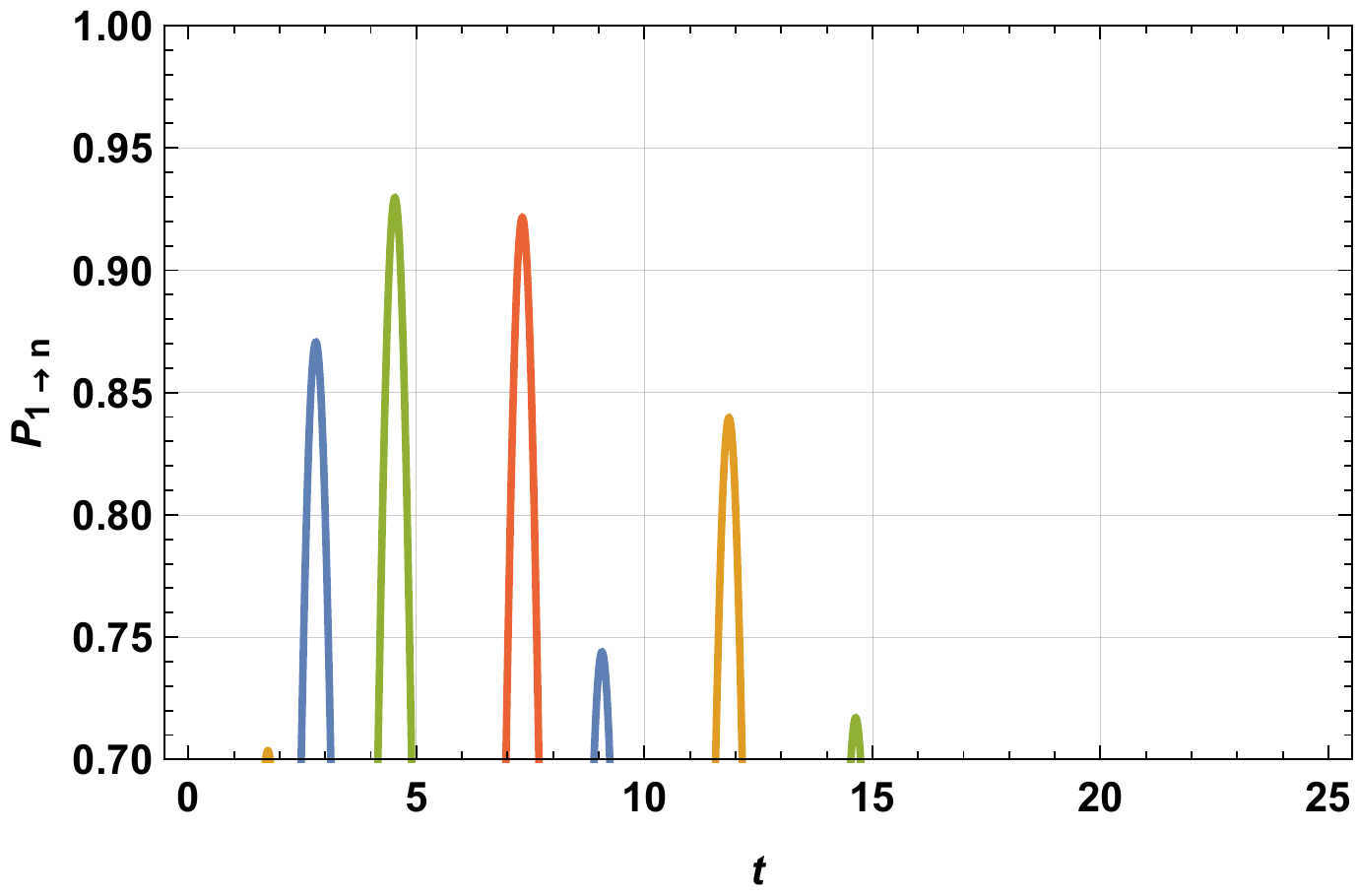}}\hfill
\subfigure[d]{\includegraphics[scale=0.59]{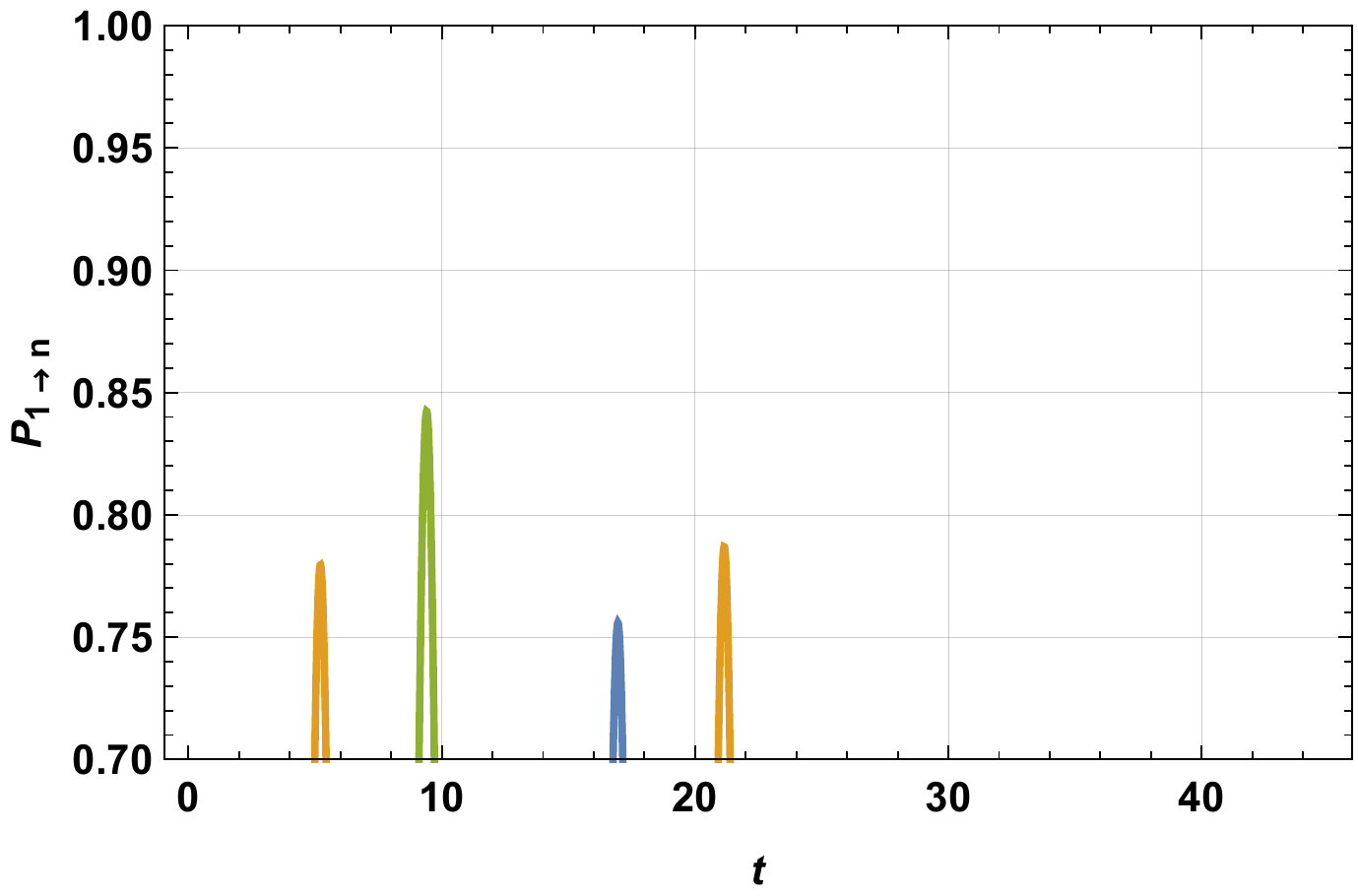}}
\caption{\small Transition probability from initial site $1$ to site $n$ on a 5-cycle ((a) and (c)) and on a 7-cycle ((b) and (d) ) as functions of time, for the resonant phase values ($\frac{\pi}{10}$ for the 5-cycle (a) and $\frac{\pi}{14}$ for the 7-cycle (b) ) and for slightly off-resonant phase values ($0.29$ for the 5-cycle (c) and $0.21$ for the 7-cycle (d) ). Different colors represent different target sites: $n=2$ in light blue, $n=3$ in orange, $n=4$ in green and $n=5$ in red. }
\label{fig:2} 
\end{figure*}

When the cycle is odd, the conclusion can change dramatically. Here there is no opposite vertex to the starting one and, by the previous argument, the non-chiral QW ($\theta = 0$) can never localize on a site different from the initial one with probability greater than $\frac12$. However, almost every $\theta \neq 0$ breaks the reflection symmetry with respect to the starting point, opening the possibility for enhanced quantum site-to-site transport. In Fig.\ref{fig:2} we plotted the transition probabilities from site $1$ to various other sites of a 5-cycle and a 7-cycle for values of $\theta = \frac{\pi}{10}, \frac{\pi}{14}$, respectively ((a) and (b)). For any other value of the phase in the relevant range $[0, \frac{2 \pi}{N} ]$, in each case the highest peaks are considerably lowered. It seems that the value $\frac{\pi}{2N}$ (where $N = 5,7$ respectively) is \emph{resonant}, although the pattern is very disordered and small changes in the precise value of the phase induce considerable shifts in the peaks at long times, which is a signature of chaos. This is illustrated by Fig.\ref{fig:2} (c) and (d), where we displayed the same set of transition probabilities for the same cycles, but with a slightly off-resonant value of $\theta$ ($\theta = 0.29$ for the 5-cycle to be compared with the resonant value of $\frac{\pi}{10} \simeq 0.31$ and $\theta = 0.21$ for the 7-cycle to be compared with the resonant value of $\frac{\pi}{14} \simeq 0.22$ ). We stress that all these peaks displayed in the plots for odd cycles are entirely due to the introduction of the phase degree of freedom, since \emph{non-chiral} CTQW on odd cycles starting in a localized state will never be found with probability higher than $\frac12$ on other sites.

We will now see how the quantum-classical distance can capture these results for the prototypical example of cycle graphs and how it can indicate the optimal phase, putting order to the opaque relationship between the values of $\theta$ and the transition probabilities. A reasonably compact expression for the quantum-classical distance at time $t$ as a function of $\theta$ can be derived from Eq.(\ref{eq:pjkring}) and the standard formulas for a continuous-time classical random walk on cycles: 
\begin{equation}
    \label{eq:DQCring}
  \begin{aligned}
   &  \mathcal{D}_{QC} (t ; \theta ) \ := \  1 \ - \ \frac{e^{-2t}}{N^2} \times \sum_{k,s=1}^{N} \exp \left[  2 t \cos \frac{ 2 \pi k}{N} \ + \right. \\
   &  \left.   - \ 4 i t \sin \left( \theta + \frac{ \pi (2s + k)}{N} \right) \sin \frac{ \pi k}{N}  \right].
    \end{aligned}
\end{equation}

\par
\begin{figure*}[!ht]
\centering
\subfigure{\includegraphics[scale=0.82]{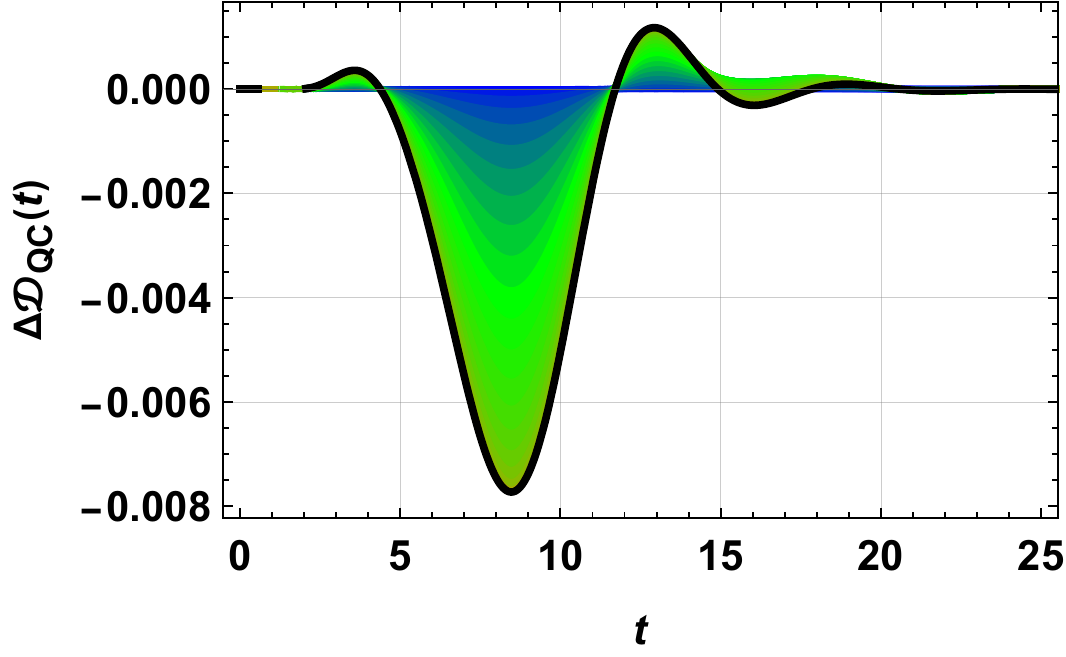}}\hfill
\subfigure{\includegraphics[scale=0.77]{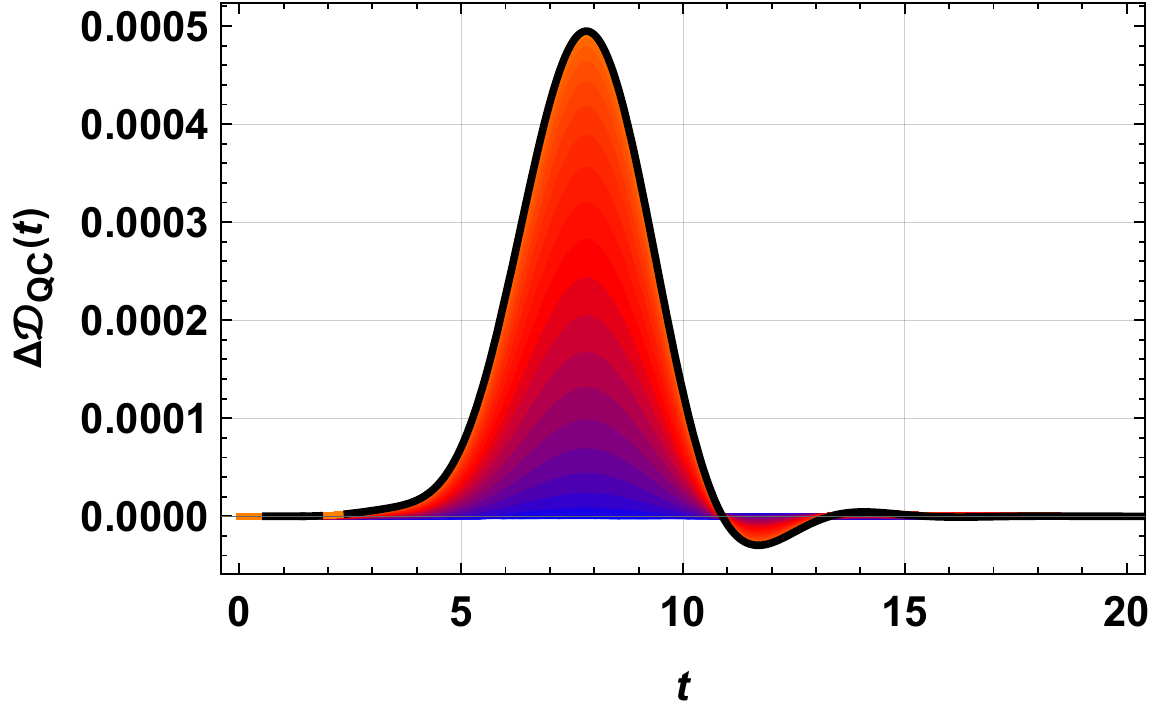}}
\caption{\small  Gain in quantum-classical distance $ \Delta \mathcal{D}_{QC}(t)$ vs. time for a 10-cycle (left) and a 7-cycle (right), with respect to the quantum-classical distance for the standard evolutions generated by the Laplacian (blue baseline in both plots). For the 10-cycle phases are increasing in the relevant range $[0, \frac{\pi}{10}]$ with a corresponding color hue from blue to green. For the 7-cycle the relevant range of phases is $[0, \frac{\pi}{14}]$ and $\theta$ increases from blue to red hues. Notice that the maximum in time of $\mathcal{D}_{QC}$ is not necessarily related with the time at which transport occurs, because the former quantity is also influenced by the time scale of the classical evolution.    }
\label{fig:3} 
\end{figure*}

\par
\begin{figure*}[!ht]
\centering
\subfigure{\includegraphics[scale=0.58]{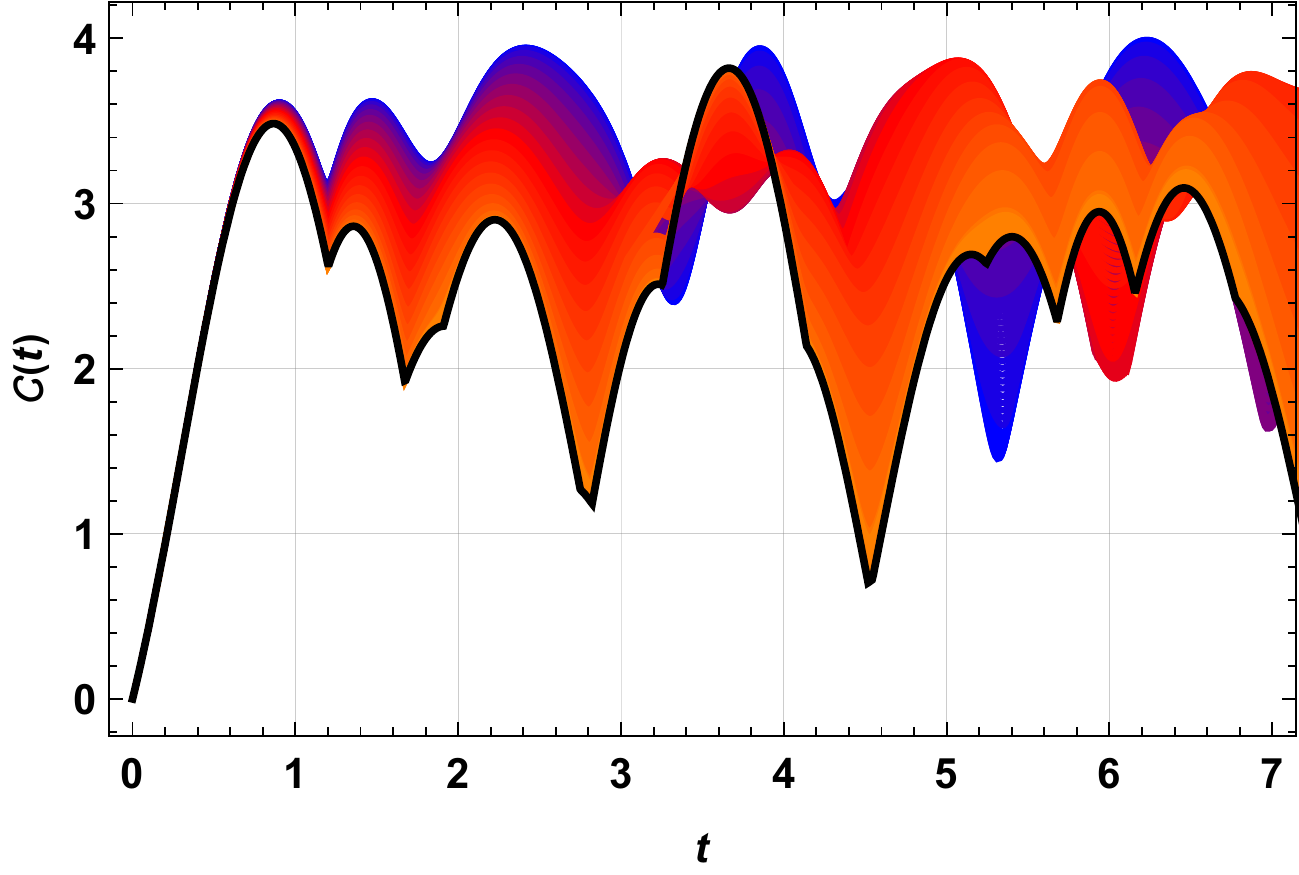}}\hfill
\subfigure{\includegraphics[scale=0.64]{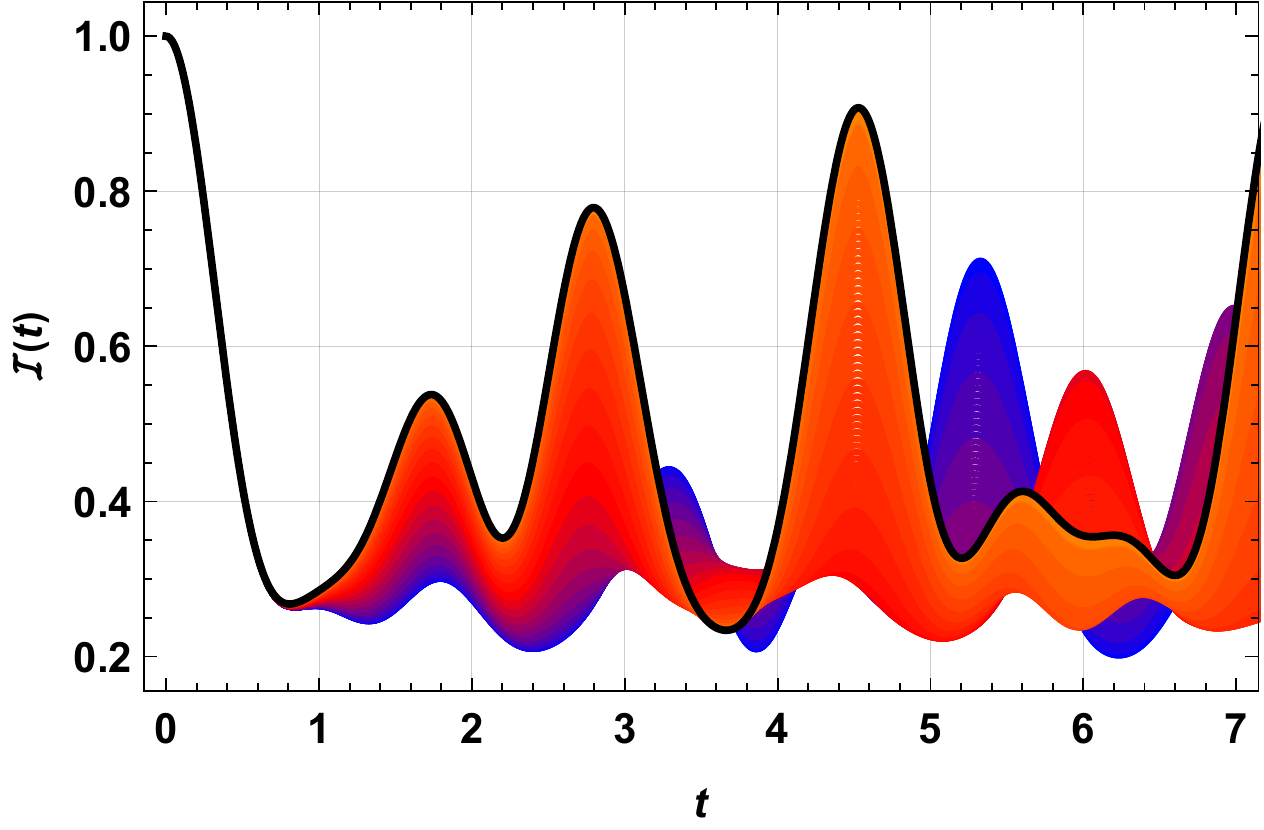}}
\caption{\small Coherence and IPR vs. time for a 5-cycle and for different values of $\theta \in [0, \frac{\pi}{10})$, corresponding to color shades from blue to orange. The black line represents the resonant phase $\frac{\pi}{10}$.   }
\label{fig:4} 
\end{figure*}

Since the dynamics is the same independently of the starting vertex, the expression for the quantum-classical distance can be computed for any localized initial state $ \vert j \rangle$ and maximization over $j$ is not necessary. It is a simple task to show that 
\[ \mathcal{D}_{QC} \left(t ; \frac{\pi}{N} + \phi \right) = \mathcal{D}_{QC} \left(t; \frac{\pi}{N} - \phi\right)\]
and, when $N$ is odd, also 
\[ \mathcal{D}_{QC} \left(t ; \frac{\pi}{2N} + \phi \right) = \mathcal{D}_{QC} \left(t; \frac{\pi}{2N} - \phi\right). \]
Therefore, when $N$ is even $\mathcal{D}_{QC}(t; \theta)$ attains all its possible values, at fixed $t$, for $\theta$ in the range $[0, \frac{\pi}{N}]$ and when $N$ is odd for $\theta$ in the range $[0, \frac{\pi}{2N}]$. 
Already for modestly sized polygons ($N \geq 7$), the variations of $\mathcal{D}_{QC}(t)$ with $\theta$ are very small. Indeed one can expand $\mathcal{D}_{QC} (t)$ around $t=0$ and notice that all terms up to order $N-1$ in $t$ when $N$ is even, and up to order $2N-1$ in $t$ when $N$ is odd, are independent of $\theta$, while an oscillatory $\theta$ dependence starts at higher orders. Because of this feeble $\theta$ dependence, we considered the difference between the quantum-classical distance with phase $\theta$ and the same quantity with zero phase, i.e. $ \Delta \mathcal{D}_{QC}(t;\theta) = \mathcal{D}_{QC} (t ; \theta ) - \mathcal{D}_{QC} (t ; 0)$. As can be appreciated in Fig.\ref{fig:3}, this quantity brings a clear order to the irregular behaviour that was seen in the transition probabilities.

This ordering is also insightful when one considers the following: for even cycles, $\mathcal{D}_{QC} (t; \theta) - \mathcal{D}_{QC} (t; 0)$ has a negative, significant dip at approximately the same time for all values of $\theta$, and it is the lowest at $\theta = \frac{\pi}{N}$, the phase which fully suppresses transport to the opposite vertex; conversely, for odd cycles we find that the same quantity \emph{peaks} at approximately the same time, with the highest peak for $\theta = \frac{\pi}{2N}$, again the resonant phase, which optimizes transport for odd cycles. Moreover, the heights of the peaks are monotonically increasing in absolute value with $\theta \in [0, \frac{\pi}{2 N} ) $, while the depths of the dips for even $N$ are monotonically increasing with $\theta \in [0,  \frac{\pi}{N} ) $. Therefore we see that finding the global maximum of $\Delta \mathcal{D}_{QC}$, with respect to $\theta$ and $t$, at a maximum (resp. minimum) of $\mathcal{D}_{QC}(t; \theta)$ in time, correctly spots the least classical (resp. the nearest to classical) evolution among all possible chiral CTQW on the same cycle. 
Intuitively, we can understand the reason of its effectiveness: $\mathcal{D}_{QC}(t)$ is maximized for the quantum evolution that departs the most from the classical one, the latter being slowly and diffusively
spreading towards the homogeneous distribution. Therefore, a chiral CTQW which evolves to nearly localized states and could be optimal for transport, will also maximizes $\mathcal{D}_{QC}(t)$. To further support this claim, we can look at the phase dependence of the functions $\mathcal{C}(t)$ and $\mathcal{I}(t)$, as in Fig.\ref{fig:4} for a 5-cycle. Again, the resonant phase (black line, $\theta = \frac{\pi}{10}$) is associated with higher localization (lower coherence and higher IPR values) especially at short times, and also the overall trend seen for the quantum-classical distance is respected at least at short times, with values of $\theta$ close to $0$ leading to the opposite behavior. We conclude this Section by noting that some results about so-called \emph{pretty good universal transport} for chiral CTQW on cycles with a prime number of vertices are known in the mathematical literature \cite{mathcycles}, although they have little role in a physical context since no bound on the time needed for transport is considered there.

\vspace{5mm}
\par
\section{Complete graphs}
\label{sec:V}
Having discussed cycle graphs characterized by minimal connectivity, we will now examine the regular graphs with maximal connectivity, i.e. complete graphs. In a complete graph just a minority of phases of $\mathrm{H}$ can be ignored by gauge invariance, and we expect that the role played by the new degrees of freedom will be major for this topology.

At this point it is worth mentioning that for standard quantum walks with real Hamiltonian generators, the dynamics of the walker on a complete graph with localized starting condition is equivalent to the dynamics on a star graph with the same number of vertices, when starting at the core vertex \cite{Xu_2009}. In other words, if the Hamiltonian is real, many links of the complete graph can be eliminated without changing the evolution. This is intuitive by symmetry arguments: since all vertices are connected to the initial one, the amplitudes in all the vertices except the first will be equal at all times and probability will never flow through links that connect these vertices, therefore they are irrelevant for the dynamics. Interestingly, the addition of phases radically changes this conclusion since, as we shall see, search to the quantum speed limit and without an oracle can be achieved on a complete graph with appropriately chosen phases, while this is clearly impossible with the star graph which has no non-trivial phase degrees of freedom, being a tree graph. The upshot is that the generalization to Hermitian Hamiltonians and chiral CTQWs is more powerful than previously imagined, because it explicitly differentiates between graph topologies that, for some initial conditions, would be completely equivalent for evolutions generated by the simple Laplacians.\\

\par
\subsection{Random chiral Hamiltonians for complete graphs}
\begin{figure*}[!ht]
\centering
\subfigure{\includegraphics[scale=0.65]{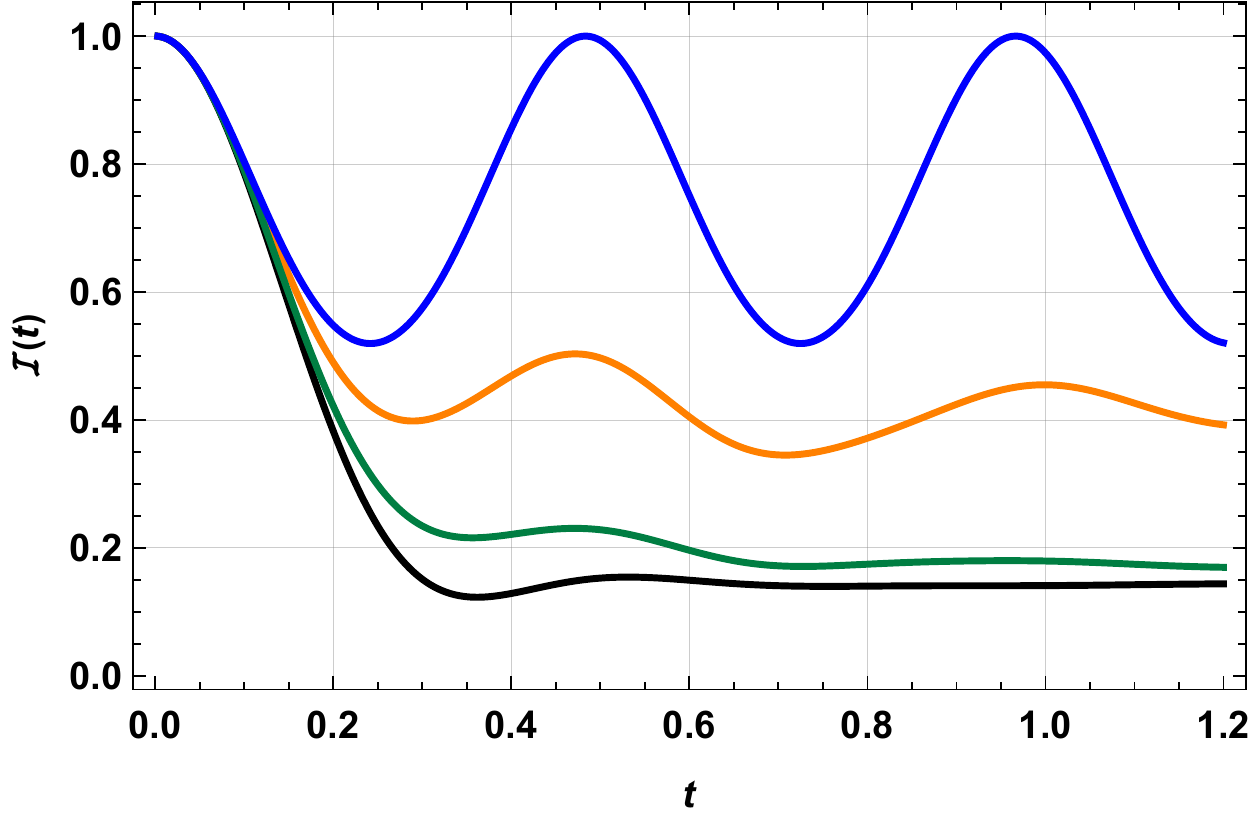}}\hfill
\subfigure{\includegraphics[scale=0.65]{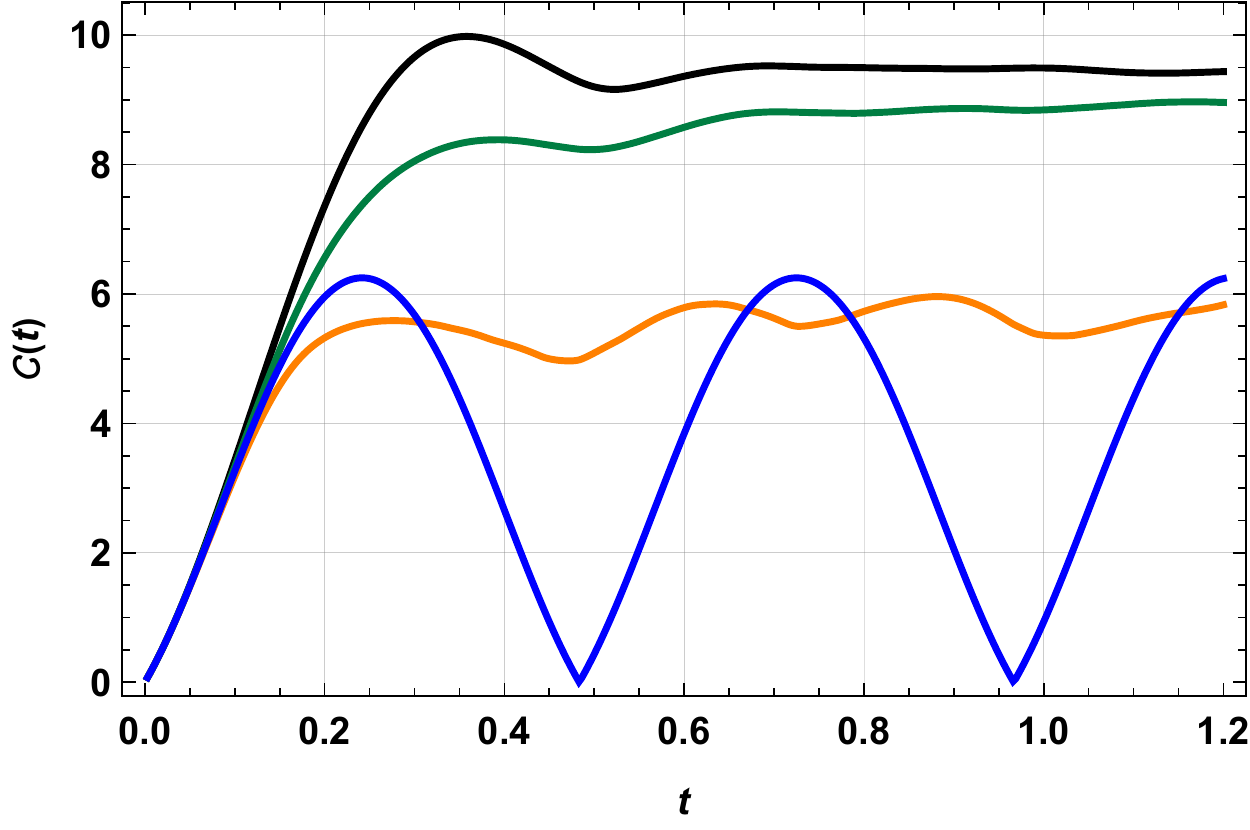}}\hfill
\caption{\small  IPR $\mathcal{I}(t)$ and coherence $\mathcal{C}(t)$ vs. time of a continuous-time quantum walker on a complete graph with $N=13$ sites with localized initial condition. The standard evolution generated by $\mathrm{H} = \mathrm{L}$ is depicted by the blue curves. The orange curves are averages over $400$ Hamiltonians with a single, randomly generated phase attached to all links in the "positive" direction. The dark-green curves resulted from the average of $400$ Hamiltonians with two, independent randomly generated phase randomly attributed to each link in positive direction. Black curves correspond to the random assignment of independent phases to each link, still averaged over $400$ runs. }
\label{fig:5} 
\end{figure*}

We initially investigated the effect of randomly added phases on quantum-classical distance, 1-norm of coherence and IPR. Because of the large number of free parameters, we adopted different strategies to better explore the parameters space. In Fig.\ref{fig:5} a comparison between the averaged coherence and IPR for different phase choices on the complete graph with $N=13$ is shown. The blue curves corresponds to the non-chiral choice with $\mathrm{H}= \mathrm{L}$. The orange curves are averages over 400 configurations where a single phase $e^{i \phi}$ is generated randomly and then attached to all the links of the complete graph in the direction $j \to k$ for $k > j$. 
We infer that a typical phase different from $\phi = 0$ attached to each link increases the average coherence and decreases the IPR with respect to $\mathrm{H} = \mathrm{L}$. 
 If, instead, we stochastically distribute two random phases $e^{i \phi_{1} }, e^{i \phi_{2}}$ among the edges in the given direction, there is an even greater increase in the coherence and decrease in the IPR (dark green curves, again resulting from an average of 400 random configurations). Finally, if  an independent, randomly generated phase is attached to each link, the resulting average behaviour is described by the black curves. Notice that now the order is irrelevant, since all orders will be explored if the phase of each link is independent and sampled in the full range $[0,2\pi)$. Clearly this rule entails all the previous ones, but the \emph{typical} configuration contributing to the black curve will be one in which there is no correlation between phases on different links and there is essentially full \emph{disorder} in the phase degrees of freedom of the Hamiltonian. Now the IPR stabilizes to the lowest values between the examined ones, while the coherence is maximal with respect to the previous cases. These two quantities therefore hints at a \emph{phase-disorder-induced delocalization} in the evolution of a localized state on a complete graph \footnote{Here by \emph{phase disorder} we mean the randomness in the choice of phases for each link, much like the disorder of the on-site potential in Anderson localization, and \emph{not} the disorder induced by uncertainty on the values of the Hamiltonian's parameters.}. These results were checked also for $N=16$ and $N=17$ and appear robust irrespective of the number of sites.

\begin{figure}[!ht]
\centering
\includegraphics[scale=0.67]{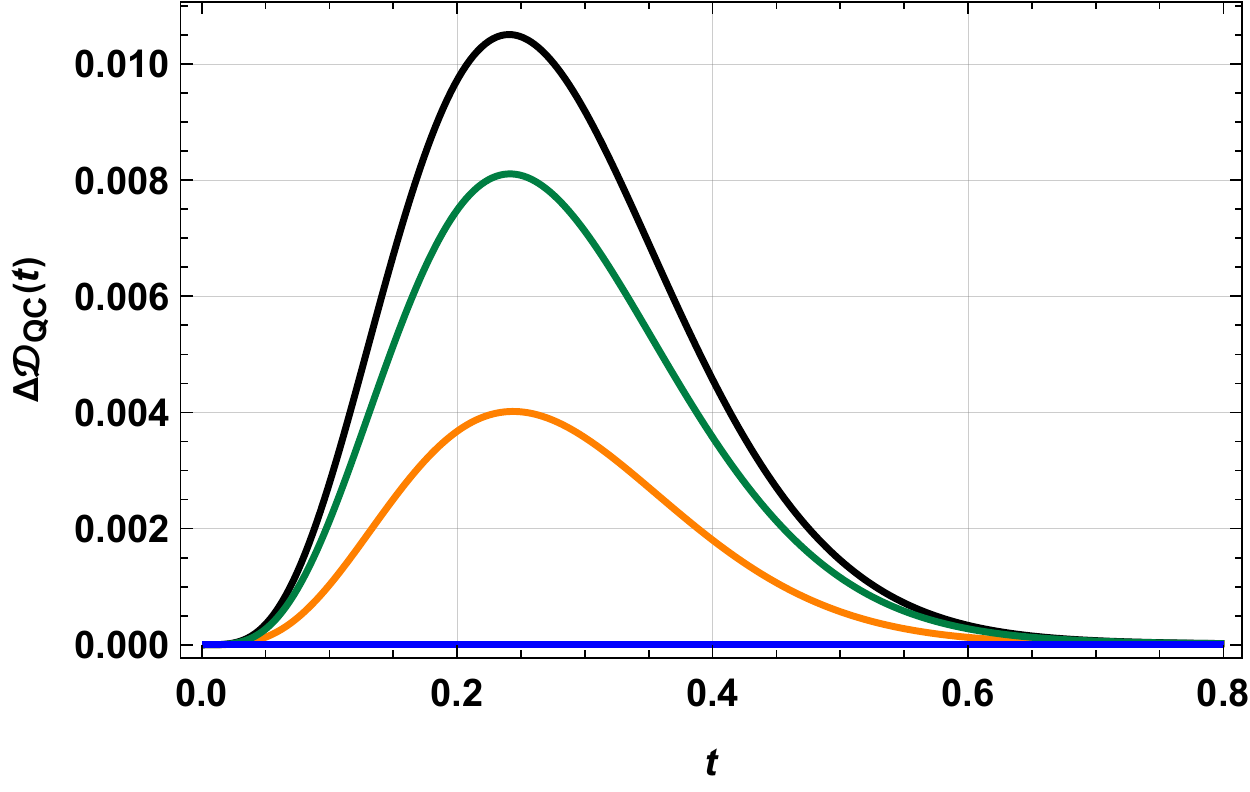}
\caption{\small Difference $\Delta \mathcal{D}_{QC}(t)$ between the quantum-classical distances of the various, averaged chiral evolutions and the non-chiral $\mathrm{H} = \mathrm{L}$ one, for a complete graph with $N=13$ sites and localized initial condition of the CTQW. Color code and meaning as for Fig.\ref{fig:5}. The reference case with $\mathrm{H}= \mathrm{L}$ is the base line in blue.   }
\label{fig:7} 
\end{figure}

Let us now look at the difference $\Delta \mathcal{D}_{QC}$ between the quantum-classical distance for these different averaged evolutions and the quantum-classical distance for the reference case $\mathrm{H}= \mathrm{L}$ (Fig.\ref{fig:7}, same color code). We see a glaring relation between this quantity and IPR or coherence: any addition of non-zero phase seems to add to the quantum-classical distance (at least \emph{on average}), and the increase goes along with the higher delocalization, as signaled by large values of the coherence and low values of the IPR.
Mind that, strictly speaking, we are considering the quantum-classical distance \emph{at fixed initial state $\vert j \rangle$}, since the full $\mathcal{D}_{QC}(t)$ would require maximization over all initial states at each time, once $\mathrm{H}$ and its phases have been specified. A crucial point here is the correlation between $\Delta \mathcal{D}_{QC}$ and localization (high values of $\mathcal{I}$ and low values of $\mathcal{C}$), which is opposite with respect to the relation found for cycle graphs.
Here those dynamics leading to less localized states are more quantum, according to the quantum-classical distance. This fact could be partially predicted from the short-time expansion of Eq.(\ref{eq:DQCshortT}) which is not monotonic in the connectivity $d_{j}$, unlike the expansions for $\mathcal{C}(t)$ and $\mathcal{I}(t)$ to the same order in $t$. This is another evidence that $\mathcal{D}_{QC}(t)$ stands out against other dynamical quantities in the successful recognition of optimal chiral quantum evolutions. Of course, the relevant task is graph-dependent, being either an hitting-type or a mixing-type task, but remarkably the quantum-classical distance seems able to optimize the parameters in both scenarios.

\subsection{Optimization of the quantum-classical distance}

Guided by these observations, the next natural step is to maximize $\mathcal{D}_{QC}(t)$ over all possible phases degrees of freedom in $\mathrm{H}$ for the complete graph. 
This has to be performed at a fixed time, that we chose afterwards by identifying the time at which $\mathcal{D}_{QC}$ attains its maximum;
however, it seems that the increasing order of $\mathcal{D}_{QC}$ curves is time independent for complete graphs (as suggested by Fig.\ref{fig:7}) and the result is the same if another time is set for the numerical maximization.  Starting with random guesses, the optimization converges to different Hamiltonians each time. However, a common feature of these optimal matrices is easy to spot: 
\begin{enumerate}
    \item \label{cond:1} The first column of an optimal $\mathrm{H}_{O}$ on the complete graph of $N$ sites is orthogonal, with respect to the Hermitian product on $\complex^{N}$, to all the rows of $\mathrm{H}$ except the first one (assuming that all the diagonal elements have been fixed to $0$, without loss of generality)
\end{enumerate}
The first column is singled out because of the choice of $\vert 1 \rangle$ as the (arbitrary) initial condition. Let us call $\underline{h}$ the first column of an optimal Hamiltonian $\mathrm{H}_{O}$. Because of the topology and the choice of diagonal elements, its entries will be:
\begin{equation}
    h_{1} = 0 \ , \ \ \ \ h_{j} = e^{i \phi_{j}} \ \ \forall j = 2,...,N
\end{equation}
and, since $\mathrm{H}_{O}$ is Hermitian, its first row will be ${ ( \underline{h}^{*} )}^{T}$, so that the scalar product between the first row and the first column is $N-1$. Let us denote by $\underline{e}_{1}$ the localized state $\vert 1 \rangle$ in matrix notation, which is the first vector of the canonical basis. Then we have:
\begin{equation}
\begin{aligned}
    \mathrm{H}_{O}^{2n} \cdot \underline{e}_{1} \ & = \ (N-1)^{n} \underline{e}_{1} \\
     \mathrm{H}_{O}^{2n+1} \cdot \underline{e}_{1} \ &= \ (N-1)^{n} \ \mathrm{H}_{O} \cdot  \underline{e}_{1} \ = \ (N-1)^{n} \underline{h} 
\end{aligned}
\end{equation}
The evolution of the initial state localized at site $1$ then follows immediately:
\begin{equation}
\label{eq:evcomplOpt}
\begin{aligned}
    & e^{- i t \mathrm{H}_{O}} \cdot \underline{e}_{1}  \ =  \\ 
    & = \ \cos [\sqrt{N-1} t  ] \underline{e}_{1}   -  \frac{i}{\sqrt{N-1}}  \sin [ \sqrt{N-1} t ] \underline{h}
\end{aligned}
\end{equation}
Notice that $\underline{h}$, as a vector of amplitudes, represents a state which is balanced between all sites except the first, to which it is orthogonal. Therefore, the evolution of Eq.(\ref{eq:evcomplOpt}) is a cyclic rotation between the initial state localized at site $1$ and the equally spread state over all other sites except the initial one, necessarily passing through an intermediate flat state:
\begin{equation}
    \vert f \rangle \ = \ \frac{1}{\sqrt{N} } \Big(    \vert 1 \rangle  \ + \ \sum_{j=2}^{N} e^{i \phi_{j}} \vert j \rangle \Big)
\end{equation}
where the amplitude to find the walker in any site of the graph is equal to $N^{-1/2}$. The time needed to reach $\vert f \rangle$ for the first time is given by:
\begin{equation}
\label{eq:tf}
    t_{f} \ = \ \frac{1}{\sqrt{N-1}} \arccos \frac{1}{\sqrt{N}} 
\end{equation}
while the time needed to reach the state $\vert \underline{h} \rangle = \frac{1}{\sqrt{N-1}} \sum_{j=2}^{N} h_{j} \vert j \rangle $ associated with the vector $\underline{h}$ and orthogonal to the initial state $\vert 1 \rangle$ is:
\begin{equation}
t_{h} \ = \ \frac{ \pi }{2 \sqrt{N-1}}.
\end{equation}
It should be emphasized that $\vert f \rangle$, as for any flat state, has maximal coherence value of $N-1$ and minimal IPR value of $\frac{1}{N}$.
Exploiting the simple structure of these optimal evolutions, an exact expression for their quantum-classical distance can also be derived:
\begin{equation}
    \mathcal{D}_{QC}^{O} (t)  =  1 - \frac{1 - e^{-N t} }{N}  - e^{-N t}   \frac{1 + \cos [ 2 \sqrt{N-1} t]}{2}.
\end{equation}

\subsection{Search to the quantum speed limit without an oracle}
Another interesting property of the "optimal" evolution described by Eq.(\ref{eq:evcomplOpt}) can be appreciated if we choose $\vert \underline{h} \rangle \ = \ \frac{i}{\sqrt{N-1}} \sum_{j=1}^{N} \vert j \rangle$, i.e. all the phases of the first column of $\mathrm{H}_{O}$ equal to $i$. This can always be achieved by an appropriate gauge transformation on any Hamiltonian that already fulfills Condition \ref{cond:1}. In this case, $\vert f \rangle = \frac{1}{\sqrt{N}} \sum_{j=1}^{N} \vert j \rangle$ is the flat state with relative phases all equal, therefore the backwards evolution from $\vert f \rangle$ is a \emph{solution to the search problem} with target vertex $1$, starting from the unbiased state and without an oracle (but with biased phases), in a time $t_{f}$, i.e. $ e^{ i \mathrm{H}_{O} t_{f} } \vert f \rangle  =  \vert 1 \rangle $. 

\begin{figure}[!ht]
\centering
\includegraphics[scale=0.8]{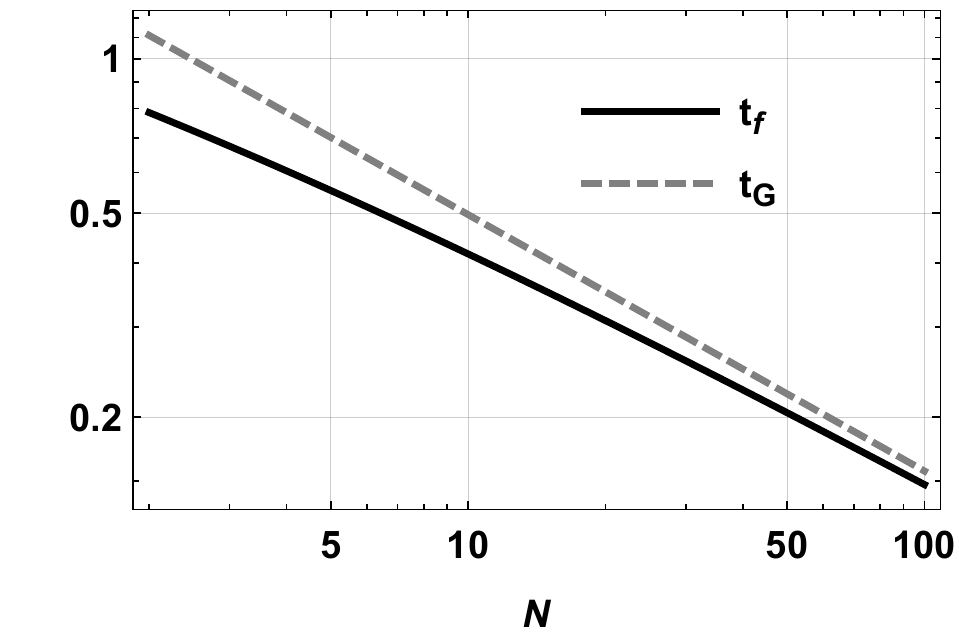}
\caption{\small Optimal search time exploiting phases $t_{f}$ (black line) and Grover's time $t_{G}$ as functions of the number of sites $N$, with log scale on both axes. $t_{f}$ is always smaller than $t_{G}$, and it is also equal to the quantum speed limit $\tau_{QSL}$  }
\label{fig:8} 
\end{figure}

Since coherence and IPR for a localized initial condition are gauge invariant quantities, the blue curves in Fig.\ref{fig:5} imply that it is not possible to reach a flat state from a localized one with any $\mathrm{H}$ which is gauge-equivalent to $\mathrm{L}$. Thus, Condition \ref{cond:1} and also our result for the search problem \emph{require} a \emph{nontrivial configuration} of the phase degrees of freedom. It is interesting to compare our search time $t_{f}$ with Grover's time and with the quantum speed limit for this evolution. Grover's time $t_{g}$ is the time required for the Grover's Hamiltonian $\mathrm{H}_{G} = \mathrm{L} - N \vert 1 \rangle \langle 1 \vert$ to search for the state $\vert 1 \rangle$ when starting from the flat state $\vert f \rangle$. The operator $N \vert 1 \rangle \langle 1 \vert$, called the \emph{oracle}, is needed to break the symmetry between all the vertices of the complete graph and to guide the evolution towards the target vertex. For a complete graph with $N$ vertices, Grover's time is:
\begin{equation}
    t_{g} \ = \ \frac{\pi}{2 \sqrt{N} } .
\end{equation}
For any $N>2$ one has $t_f < t_{g} < t_h$, therefore our search Hamiltonian is \emph{faster} than Grover's one, and does not require an oracle.
Of course, in order for this result to hold, a bias has to be present in the \emph{phases} of our optimal $\mathrm{H}$ so that the target vertex can be singled out during the evolution. This bias is in fact embodied by Condition \ref{cond:1}. 
It is now meaningful to ask whether this time $t_{f}$ reaches the ultimate time bound allowed by quantum mechanics for the evolution between these initial and final state.
To this end, we briefly recall the notion of quantum speed limits \cite{QSL03a,QSL03,QSL04,QSL17}. Consider two states $\vert a\rangle, \vert b \rangle$ on an Hilbert space $\mathcal{H} \sim \complex^{N}$ that have the \emph{same} average energy \footnote{This is clearly a necessary condition in order for the unitary evolution generated by $\hat{H}$ to bring $\vert a \rangle$ to $\vert b \rangle$. } with respect to a time-independent Hamiltonian operator $\hat{H}$ on $\mathcal{H}$. Then the \emph{quantum speed limit} $\tau_{QSL}$ is a lower bound on the time needed for the unitary evolution $e^{-i t \hat{H}}$ to rotate $\vert a \rangle$ to $\vert b \rangle$, and it is provided by the following expression:
\begin{equation}
\label{eq:QSL1}
    \tau_{QSL}   :=  \max \left\{ \frac{ \arccos \vert \langle b \vert a \rangle \vert}{ \Delta \hat{H}}  \ , \ \frac{2  ( \arccos \vert \langle b \vert a \rangle \vert )^{2} }{ \pi ( \langle \hat{H} \rangle - E_{0} ) }        \right\} 
\end{equation}
where $\Delta\hat{H} = \sqrt{\langle \hat{H}^{2} \rangle - \langle \hat{H} \rangle^{2}}$ is the standard deviation of energy of the states ( $\vert a \rangle$ or $\vert b \rangle$ ), $\langle \hat{H} \rangle$ their average energy and $E_{0}$ is the ground state's energy. 
The first quantity inside the $\max$ of the quantum speed limit can be readily computed in our case, for $\vert a \rangle = \vert f \rangle$ and $\vert b \rangle = \vert 1 \rangle$; indeed, the variance $\Delta \hat{H}$ does not depend on the phases and it is equal to $\sqrt{N-1}$ for any Hamiltonian with constant diagonal terms and compatible with the complete graph's topology. Therefore:
\begin{equation}
\label{eq:QSL2} \frac{ \arccos \vert \langle 1 \vert f \rangle \vert  }{\Delta \hat{H}} \ = \ \frac{1}{\sqrt{N-1}} \arccos \frac{1}{\sqrt{N}} \ = \ t_{f}      
\end{equation}
Since we already know that $t_{f}$ cannot be smaller than $\tau_{QSL}$, the $\max$ in the definition (\ref{eq:QSL1}) of $\tau_{QSL}$ and the result of Eq.(\ref{eq:QSL2}) already imply that $\tau_{QSL} = t_{f}$. 
This was verified by computing the second quantity in Eq.(\ref{eq:QSL1}) and checking that it is always smaller than the first for $N>3$ \footnote{The check was performed numerically because $E_{0}$ depends on the phases in a non trivial way, so that the second quantity is harder to compute in general. }. It can be shown that $\tau_{QSL}$ is also the quantum speed limit for Grover's Hamiltonian. Indeed, we highlight that $\Delta \hat{H}$ attains its largest value for the complete graph, among all the topologies that connect $N$ vertices in a simple, connected graph. We conclude that our construction exploits phases to achieve quantum search between $N$ orthogonal states without an oracle and in the least possible time allowed by quantum mechanics, without altering the on-site energies at will.
We remark that the scaling behaviour $O( N^{-1/2})$ of Grover's time is known to be already the best one, and indeed our search time $t_{f}$ follows the same asymptotic scaling for large $N$. The construction that we provided, on the other hand, achieves the goal of optimizing the constant pre-factor, which is sub-optimal for Grover's algorithm. The comparison in log-scale is shown in Fig.\ref{fig:8}.

\begin{figure*}
\centering
\subfigure{\includegraphics[scale=0.533]{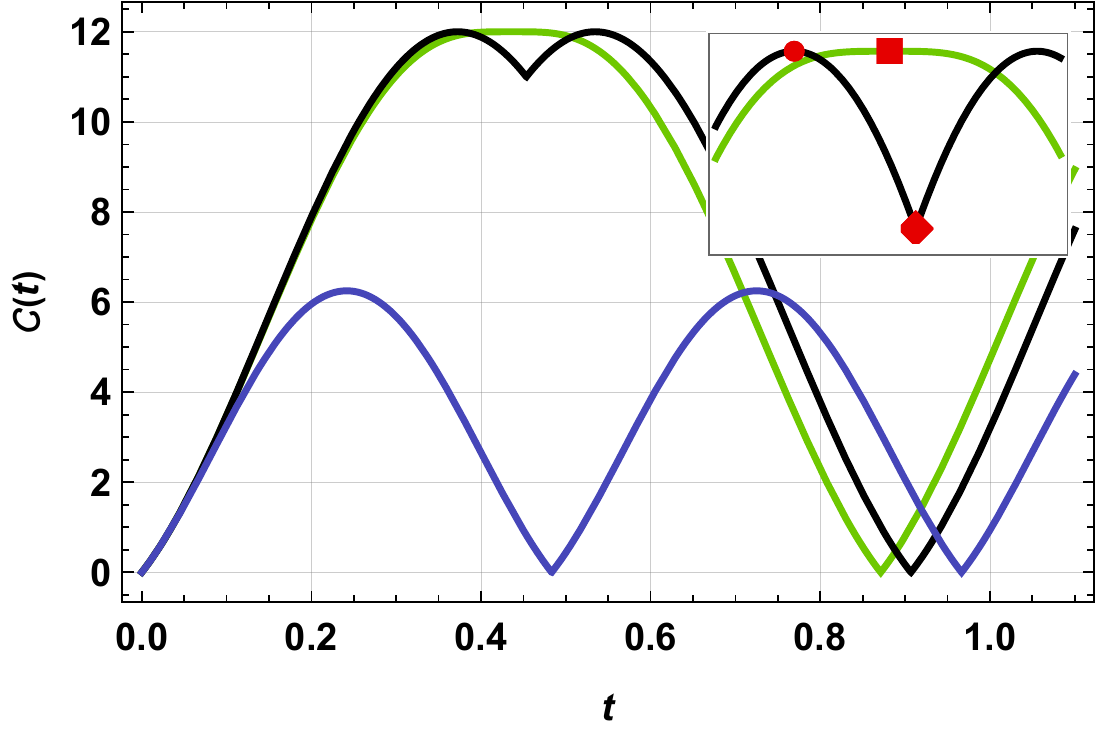}}\hfill
\subfigure{\includegraphics[scale=0.533]{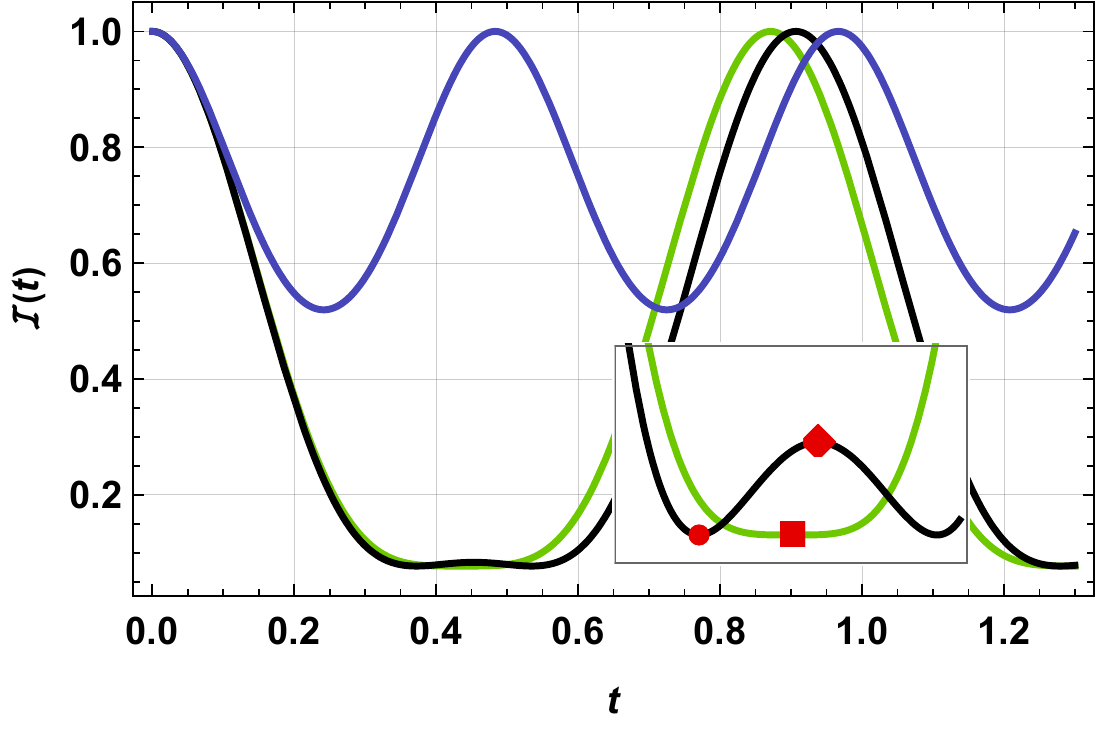}}\hfill
\subfigure{\includegraphics[scale=0.553]{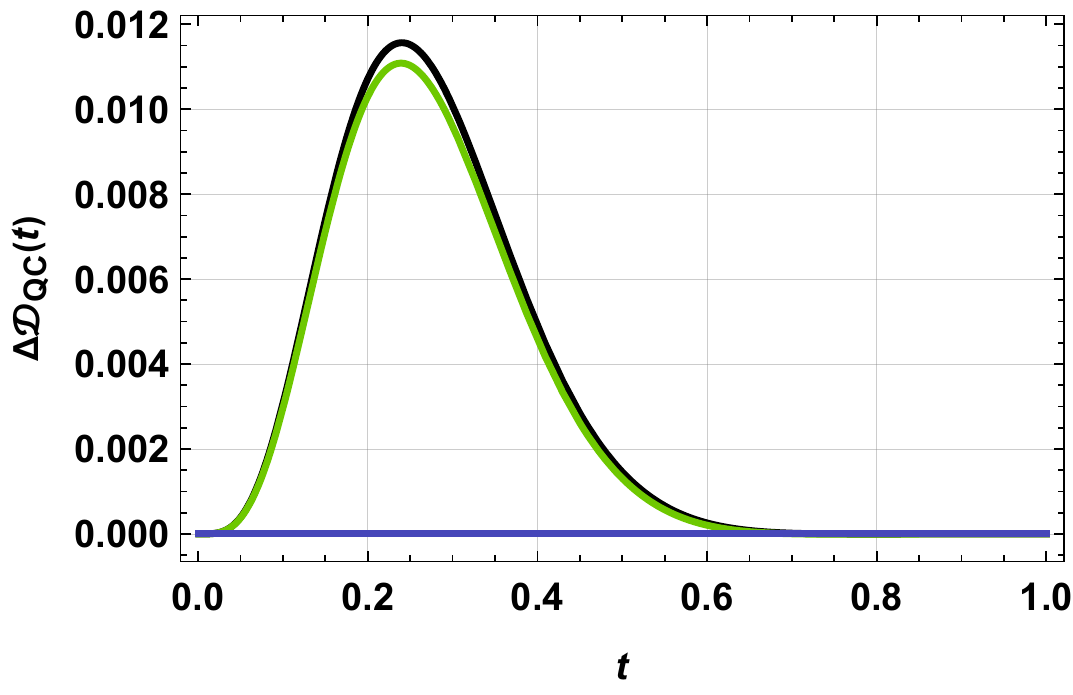}}
\caption{\small Comparison for coherence $\mathcal{C}(t)$, IPR $\mathcal{I}(t)$ and gain in quantum-classical distance $\Delta \mathcal{D}_{QC}(t)$ for evolutions generated by $\mathrm{H}_{O}$ (black curves), $\mathrm{H}_{G}$ (light-green curves) and $\mathrm{H}= \mathrm{L}$ (blue curves) for a complete graph of $N=13$ sites and starting with a localized state. The insets in plots of $\mathcal{C}(t)$ and $\mathcal{I}(t)$ show a magnification of the region for $0.3 < t < 0.5$, where times $t_{f}$ (red circle), $t_{G}$ (red square) and $t_{h}$ (rotated red square) are located.    }
\label{fig:9} 
\end{figure*}

To further illustrate the differences between the evolution induced by Grover's Hamiltonian $\mathrm{H}_{G}$ and the optimal solution $\mathrm{H}_{O}$ that we found, we plotted in Fig.\ref{fig:9} the time behavior of the quantities $\mathcal{C}(t), \mathcal{I}(t)$ and also the gain in quantum-classical distance $\Delta \mathcal{D}_{QC}$ with respect to the reference $\mathrm{H} = \mathrm{L}$, again for a complete graph with $N=13$ sites and starting at vertex $\vert 1 \rangle$. Black curves depict the evolution generated by $\mathrm{H}_{O}$, while Grover's evolution is in light-green. The blue curves correspond to the non-chiral $\mathrm{H} = \mathrm{L}$ choice. The insets show a magnification of the regions where the relevant times happen: the red circle indicates the time $t_{f}$, the red square designates time $t_{G}$ and finally the rotated square indicates $t_{h}$. 

Importantly, the gain in quantum-classical distance with respect to the non-chiral $\mathrm{H} = \mathrm{L}$ choice is also maximal for the evolution generated by $\mathrm{H}_{O}$, even when compared with Grover's evolution, a non obvious fact since the maximization was performed without considering diagonal degrees of freedom. Moreover, the optimal evolution outperforms the best evolutions found in Fig.\ref{fig:5} by randomly generating phases according to IPR and coherence, and correspondingly $\Delta \mathcal{D}_{QC}$ is in fact higher at all times. 

It remains to be shown that Condition \ref{cond:1} can indeed be fulfilled by some choice of phases for any $N$, without relying on the optimization of $\mathcal{D}_{QC}$. This is carried out in detail in Appendix \ref{apx:constr}, but it should be clear that Condition \ref{cond:1} arose solely from the maximization of the quantum-classical distance, thereby corroborating the power of this method. \\

To conclude with this class of examples, let us remark that with non-chiral CTQWs on complete graphs with $N$ vertices it is impossible to observe \emph{instantaneous uniform mixing} \cite{ahmadimixing}, where the walker completely delocalizes to a uniform superposition of all basis states at a certain instant during its evolution, except for $N = 2, 3, 4$. Our result shows that, with the generalization to \emph{chiral} quantum walks, but retaining unitarity, instantaneous uniform mixing is achievable on \emph{any} complete graph in a remarkably short time $t_{f}$ given by Eq. \eqref{eq:tf}, which is $O(1/\sqrt{N})$ for large $N$. Importantly, this is a quadratic speedup with respect to the $O(1)$ scaling of mixing on hypercube graphs \cite{mooremixing} \footnote{ In the  literature on quantum algorithms, it is common wisdom to rescale the adjacency matrix of regular graphs by their connectivity to compare evolution times. With that convention, the time needed to achieve uniform mixing with the scaled  adjacency  matrix  on hypercube graphs is $O(N)$, while the time to perform the same task on complete graphs with our \emph{chiral} protocol is $O(\sqrt{N}$). }  and it also holds for generic number of sites, whereas the hypercube protocol applies only if $N$ is a power of $2$.

\par
\section{Quantum switches}
\label{sec:VI}
\begin{figure}
\centering
\includegraphics[scale=0.5]{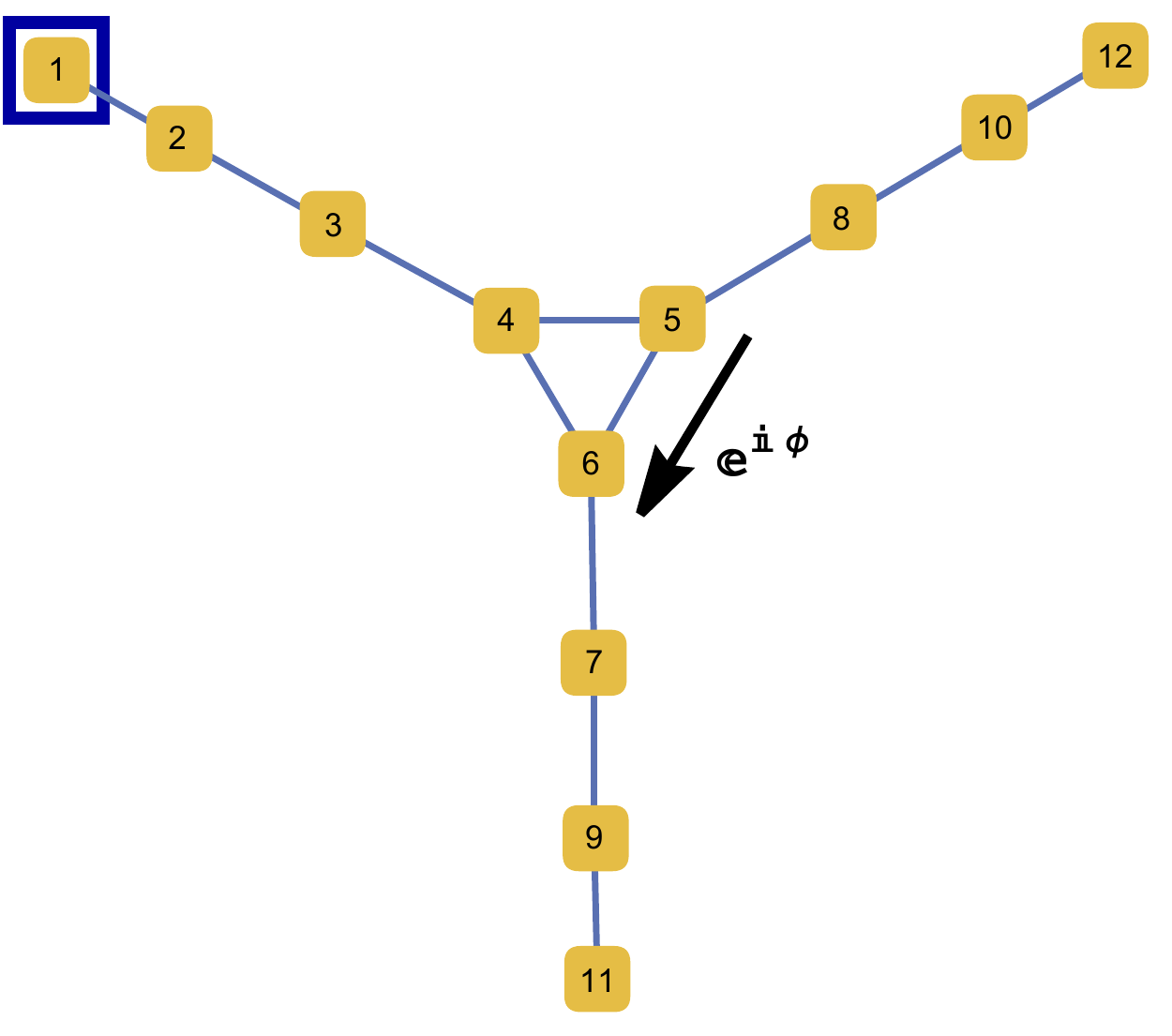}
\caption{\small Graph of the quantum switch with 12 sites. The walker starts at site $1$ (boxed in blue) and the link between site $5$ and site $6$ bears a phase of $e^{i \phi}$ in the Hamiltonian (black arrow). }
\label{fig:10} 
\end{figure}

Here we consider graphs like the one depicted in Fig.\ref{fig:10}, which are seldom referred to as \emph{quantum switches}. Since the graph is planar and there is a single loop, just one phase will affect transition probabilities between sites, and we attach it on the link which closes the triangle and is opposite to vertex $1$, calling it $e^{i \phi}$ in the direction specified in the figure. 

\begin{figure}
\centering
\includegraphics[scale=0.68]{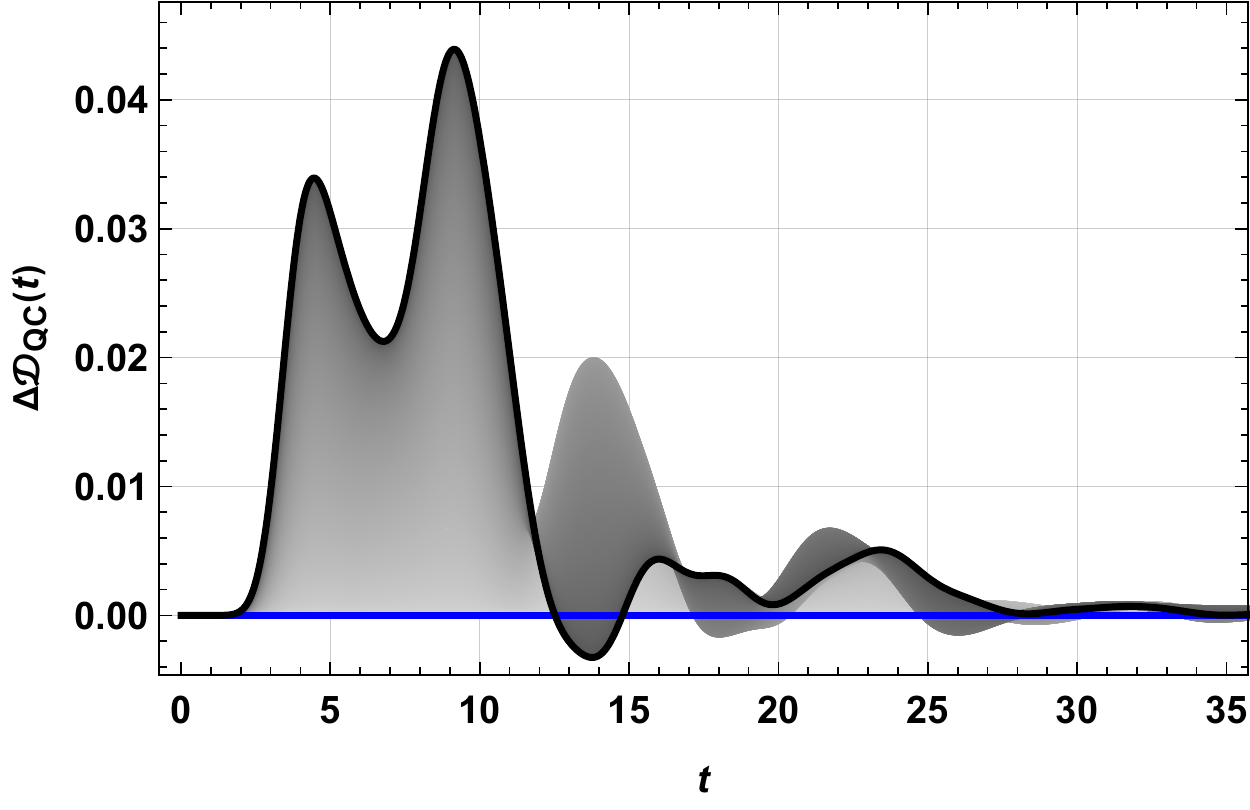}
\caption{\small Difference between $\mathcal{D}_{QC}$ for the adjacency matrix of the $12$-sites quantum switch with an adjoined phase and the same quantity for the standard adjacency matrix. The value of $\phi$ ranges from $0$ to $\frac{\pi}{2}$ for increasingly darker shades of gray. The black line results from the resonant phase $\frac{\pi}{2}$ and the blue baseline is the non-chiral evolution.}
\label{fig:11} 
\end{figure}

These graphs were considered in \cite{cqw1} as examples of the advantage provided by \emph{chiral} quantum walks over the ones defined by the Laplacian or the adjacency matrix. Indeed, it was shown that a resonant value of $\phi = \frac{\pi}{2}$ suppresses transport from vertex $1$ to vertex $11$, with reference to Fig.\ref{fig:10}, while enhancing transport from vertex $1$ to vertex $12$. Unlike all the other graphs considered so far, the quantum switch is non-regular, meaning that the connectivity is not the same for all sites and the Laplacian or the adjacency matrix of the graph generate different quantum evolutions. Here we first consider the adjacency matrix $\mathrm{A}$  for a $12$-sites switch, and then attach a phase to the link between vertex $5$ and vertex $6$. In Fig.\ref{fig:11} we again plot the difference between $\mathcal{D}_{QC}(t)$ for the Hamiltonian with such a phase and the same quantity without the phase. The value of $\phi$ is increasing between $0$ and $\frac{\pi}{2}$ from light gray to black curves, and we checked that larger values of $\phi$ are redundant.

The comparison with coherence and IPR shows a similar relation to that seen for cycle graphs: more localized evolutions, leading to higher values of IPR and lower values of coherence, are associated with higher gains in $\Delta \mathcal{D}_{QC}$. The resonant phase $\phi = \frac{\pi}{2}$ is clearly identified as the black curve which maximizes $\Delta \mathcal{D}_{QC}$. It achieves a transport fidelity of $ \sim 0.77$ to the target state $ \vert 12 \rangle$ when starting in $\vert 1 \rangle$ in a time $t \sim 5$. Surprisingly, this transport probability is considerably higher than the value that would be reached in a comparable time range for a simple chain of $8$ sites, i.e. if the third arm of the triangle (composed of vertices $6$, $7$, $9$ and $11$) were not there to begin with, assuming the Laplacian as generator in this latter case.\\

When we take the Laplacian in place of the adjacency matrix as the starting point, instead, the result differs qualitatively. Now for any choice of the free phase, localization on one branch of the switch is never as effective as the previous case. Accordingly, $\mathcal{D}_{QC}$ attains lower values now and its variations with phases are smaller, favoring the choice $\mathrm{H} = \mathrm{A}$ for quantum transport on this topology. 
This example also provides further motivation in favor of the use of the adjacency matrix in place of the Laplacian for CTQWs on finite, non-regular graphs. Indeed, while having the connectivities on the diagonal is natural if the differential form of the Laplacian has to be recovered in the continuum limit (without an additional potential field landscape), this request is not so meaningful for small, non-regular graphs that do not embody a discretization of a continuous space in any obvious way.

\par
\section{Cube graph}
\label{sec:VII}
On the cube graph, perfect transport between opposite vertices is possible with a standard CTQW \cite{childs2004spatial}. Moreover, being part of the hypercubes family, the CTQW starting from any localized state will evolve unitarily towards a maximally coherent state, exhibiting instantaneous uniform mixing \cite{mooremixing}. In other words, both goals considered in this work are already achieved by a standard, non-chiral CTQW on the cube graph. Notice also that, once the phases degrees of freedom are taken into account, instantaneous uniform mixing and search without an oracle become equivalent: if the former is possible, starting from a localized state, the final uniform state will in general encode the information about the starting vertex in the relative phases. These can be transferred to the Hamiltonian with a gauge transformation, then by reversing the sign of $\mathrm{H}$ one obtains a chiral CTQW which evolves the uniform state with no relative phases towards a localized one, effectively performing search without an oracle. However, since this is achieved just by a gauge transformation, it will not affect $\mathcal{D}_{QC}$ and it is for all purposes equivalent to the initial non-chiral CTQW. In accordance with these observations, the optimization of the quantum-classical distance over the phases degrees of freedom on the cube graph does not provide better options: the best case for quantum hitting and mixing is already the non-chiral one. On the other hand, by minimizing $\mathcal{D}_{QC}(t)$ at short times, one finds a choice of phases that completely suppresses transport to half of the vertices of the cube, more precisely to those that are not connected to the starting one.

\par
\section{Conclusion}
The opportunities stemming from a generalization of the dynamical generator of continuous-time quantum walks to a generic Hermitian matrix compatible with the graph topology have just begun to be explored and recognized in the quantum information literature. Together with the new phase degrees of freedom that permit this generalization, also comes the issue of finding the best Hamiltonian for a certain quantum task on a given graph topology. After defining continuous-time classical and (chiral) quantum walks on graphs and gauge transformations on the latter, we put at the center three gauge-invariant dynamical quantities that should help the exploration of the effects of the phase degrees of freedom on the evolution of chiral CTQW. The first two, namely the 1-norm of coherence in the on-site basis and the Inverse Participation Ratio, quantify the degree of quantum coherence and of localization of quantum walker's state at any given time, the former being relevant because it is an inherently quantum property, while the latter can spot quantum transport. The third dynamical indicator is the quantum-classical distance, which aims at gauging the difference between a (possibly chiral) quantum evolution and the unique classical one on the same graph. Comparing the short-time expansions of the three indicators, we see that there is a correlation between them, but with no unambiguous common structure. Relying on four significant examples of graphs (some of which are actually infinite graph families) we argue that the quantum-classical distance effectively spots the "optimal" chiral quantum walk from the point of view of different tasks, depending on the topology.

\par 
Cycle graph provide the first example and a testbed for the proposal, since some analytical results can be derived in this simple case. The maximization of $\mathcal{D}_{QC}(t)$ with respect to the phase correctly suggests that quantum transport on odd cycles is enhanced for a resonant value of the phase, while confirming that on even cycles the standard choice of the Laplacian is already the best one. For complete graphs, the number of relevant phases is very large, and a preliminary exploration of the parameters space by randomly generating chiral Hamiltonians suggests a clear correlation between greater values of $\mathcal{D}_{QC}$ and \emph{phase-disorder-induced delocalization}, witnessed by low values of IPR and high values of coherence when all phases are random and independent. A systematic maximization of $\mathcal{D}_{QC}(t)$ (at fixed, short times) indeed identifies a chiral quantum evolution on the complete graph which achieves maximal coherence and lowest IPR in very short times. A neat property of these optimal Hamiltonians is recognized, allowing us to show that they can be used for quantum search to the optimal quantum speed limit and without an oracle, outperforming Grover's algorithm in the constant pre-factor. \\

\par The third example is already known from the literature on chiral CTQWs as a \emph{quantum switch}. Again, maximization of $\mathcal{D}_{QC}(t)$ identifies the best phase for directional transport on these topologies. Also, since these graphs are not regular, an ambiguity between the use of the Laplacian or of the adjacency matrix arises, and we argue that the latter is better for directional transport and perhaps more natural.

\par Finally we examined the cube graph, which is known to exhibit perfect quantum transport for $\mathrm{H} = \mathrm{L}$, as part of the family of hypercubes. Numerical simulations suggest that indeed, for this topology, the standard non-chiral case has already the highest value of $\mathcal{D}_{QC}(t)$ against all other possible choices of phases, indicating that faster quantum transport could be impossible here (without ad hoc modifications of on-site energies). However, the \emph{minimal} value of the quantum-classical distance also spots an interesting chiral dynamics, where half of the vertices of the cube, namely those disjoint from the initial one, are never visited by the walker. Together with analogous conclusions for even cycles, this suggests that \emph{minimization} of $\mathcal{D}_{QC}(t)$ can identify \emph{suppression of transport} whenever this can happen. \\

\par
\begin{appendix}

\section{Explicit construction of complete graph chiral Hamiltonians for quantum search to the speed limit}
\label{apx:constr}

First notice that, for any $N \geq 4$, Condition \ref{cond:1} cannot be fulfilled by simply filling the off-diagonal entries of $\mathrm{H}$ with any combination of $\pm 1$ and zeroes on the diagonal (with the constraint of having an Hermitian matrix compatible with the complete graph's topology). One has necessarily to resort to complex numbers. When $N$ is even, a combination of $\pm i$ does the trick. Indeed, by choosing:
\begin{equation}
\begin{aligned}
    & \left[ \mathrm{H} \right]_{1j} =  i  , \ \ \ \ \ \  \left[ \mathrm{H} \right]_{jj} = \left[ \mathrm{H} \right]_{11} = 0   \ \ \  \ \forall j = 2,...,N   \\
    & \left[ \mathrm{H} \right]_{jk} = (-1)^{j+k} i \ \ \  \ \ \  \forall \  k > j > 1 \\ 
    \end{aligned}
\end{equation}
and with the Hermitian constraint $\left[ H \right]_{jk} = \left[ H \right]^{*}_{kj} \ \forall j,k = 1,..., N $. It is immediate to check that indeed, for $N$ even, the first column of $\mathrm{H}$ is orthogonal to all the rows except the first one, with which it has an inner product of $N-1$. For example, for $N=6$:
\begin{equation}
   \mathrm{H} \ = \  \left(  \begin{array}{cccccc}
         0 &  i  & i & i & i & i \\
         -i & 0  & -i & i & -i & i \\
         -i & i  & 0  & -i & i & -i \\
         -i & -i & i  & 0 &  -i & i \\
         -i & i & -i  & i &  0 & -i \\
         -i & -i & i  & -i &  i & 0 \\
    \end{array}  \right)
\end{equation}
\\

For odd $N$, we found the following construction:
\begin{equation}
    \begin{aligned}
     & \left[ \mathrm{H} \right]_{1j} = - i  , \ \ \ \ \ \  \left[ \mathrm{H} \right]_{jj} = \left[ \mathrm{H} \right]_{11} = 0   \ \ \  \ \forall j = 2,...,N   \\
    & \left[ \mathrm{H} \right]_{jk}  =  \exp  \left\{   \frac{ 2 \pi i}{N-2}  \left[  k - j +\frac{N-3}{2} \right]  \right\}  ,  \ \ \  \forall   k > j > 1 \\ 
    \end{aligned}
\end{equation}
E.g., for $N=5$:
\begin{equation}
   \mathrm{H} \ = \  \left(  \begin{array}{cccccc}
         0 &  i  & i & i & i  \\
         -i & 0  & e^{- \frac{ 2 i \pi}{3}} & 1 & e^{ \frac{2 i \pi}{3}}  \\
         -i & e^{ \frac{2 i \pi}{3}}  & 0  & e^{- 2i \frac{\pi}{3}} & 1  \\
         -i & 1 & e^{ \frac{2 i \pi}{3}}  & 0 &  e^{-  \frac{2 i \pi}{3}}  \\
         -i & e^{- \frac{ 2 i \pi}{3}} & 1  & e^{  \frac{2 i \pi}{3}} &  0  \\
    \end{array}  \right)
\end{equation}

\end{appendix}

\bibliography{DQCchiralCTQW.bib}

\end{document}